\pdfoutput=1
\documentclass[pra,aps,showpacs,unsortedaddress,superscriptaddress,twocolumn]{revtex4}
\usepackage{graphicx}
\usepackage{float}
\providecommand{\e}[1]{\ensuremath{\times 10^{#1}}}
\begin{document}
\title{Phonon mediated quantum spin simulator employing a planar ionic crystal in a Penning trap}

\author{C.-C. Joseph Wang}
\affiliation{Department of Physics, Georgetown University, Washington, DC 20057, USA}
\affiliation{Theoretical Division, Los Alamos National Laboratory, Los Alamos, NM 87545, USA }

\author{Adam C. Keith}
\affiliation{Department of Physics, Georgetown University, Washington, DC 20057, USA}
\affiliation{Department of Physics, University of Colorado at Boulder, CO 80309}

\author{J. K. Freericks}
\affiliation{Department of Physics, Georgetown University, Washington, DC 20057, USA}

\date{\today}


\def\bx{{\bf x}}
\def\bk{{\bf k}}
\def\br{{\bf r}}
\def\bu{{\bf u}}
\def\half{\frac{1}{2}}
\def\args{(\bx,t)}

\begin{abstract}
We derive the normal modes for a rotating Coulomb ion crystal in a Penning trap, quantize the motional degrees of freedom,
and illustrate how they can by driven by a spin-dependent optical dipole force to create a quantum spin simulator on a triangular lattice with hundreds of spins.
The analysis for the axial modes (oscillations perpendicular to the two-dimensional crystal plane) follow a standard normal-mode analysis,
while the remaining planar modes are more complicated to analyze because they have velocity-dependent forces in the rotating frame.
After quantizing the normal modes into phonons, we illustrate some of the different spin-spin interactions that can be generated by
entangling the motional degrees of freedom with the spin degrees of freedom via a spin-dependent optical dipole force.
In addition to the well-known power-law dependence of the spin-spin interactions when driving the axial modes blue of phonon band, we notice certain parameter regimes in which the level
of frustration between the spins can be engineered by driving the axial or planar phonon modes at different energies. These systems may allow for the analog simulation of  quantum spin glasses with large numbers of spins.
\end{abstract}
\maketitle

\section{Introduction}

Richard Feynman motivated the idea of using analog quantum simulation as a means to describe the behavior
of complex quantum-mechanical systems~\cite{feynman}. This idea has been experimentally realized for the transverse-field Ising model in cold ion traps in one-dimension~\cite{two-ion,Kim1,Kim2,Edwards,Islam}, in neutral atoms driven to a nonequilibrium state~\cite{Markus}, and with preliminary results also available for cold ion traps in two-dimensions~\cite{John}.
The tunability and control of cold atom and cold ion systems are unmatched in
traditional (real-material)  condensed matter systems and therefore they are an ideal platform for examining idealized condensed matter systems and searching for nontrivial quantum ground states. In addition, the  isolation of the system from the environment with long decoherence times (in milliseconds) also offers tremendous opportunities for the study of non-equilibrium coherent many-body dynamics in real time such as thermalization~\cite{Marcos, dynamic QS} and quench dynamics~\cite{Santos,kastner} in cold atoms.

Cold ions in traps have been proposed theoretically as potential emulators for effective spin models~\cite{spin-spin-interactions}. For the case of linear ion chains in Paul traps, the experimental protocol has been well established~\cite{Kim2,Edwards,Islam}. However, there are still difficulties in scaling up Paul-trap based systems to more than approximately sixteen ions while maintaining good quantum control, although next generation traps are planned
for up to potentially 100 ions.
Theoretically, phonon modes in linear Paul traps have been well understood for a decade~\cite{James} and are essential for the realization of the spin models in the system (although effects of micromotion on two-dimensional crystals in Paul traps are more complicated~\cite{paultrap_2d}).
On the other hand, ion crystals in Penning traps can be easily stabilized with several hundreds of ions in a two dimensional structure and can  potentially simulate quantum phases beyond what modern computers can simulate. Recently, a feasibility study has shown that it is possible to generate spin-spin interactions in these systems that decay like a tunable power law for long distances~\cite{John}.
Similar to cold ions in linear Paul traps, a  theoretical understanding of phonon modes is needed as a prerequisite for the adiabatic state evolution of these systems.

In this paper, we provide a complete theory for the oscillatory normal modes in a Penning trap, tuned to
parameters similar to those used in current experiments.  We then quantize these phonons and show how
they can be used with a spin-dependent optical dipole force to generate effective spin-spin interactions.
Recently, others have shown how to find equilibrium positions for similar trapping potentials~\cite{Chinese}, and have evaluated the phonon spectra using different methods from the ones we employ~\cite{Jake}.

The organization of the paper is as follows.
In Sec. II, we formulate the theory for the normal modes of cold ions in Penning traps.
 We determine the ground state of the crystal configuration by finding the static equilibrium of the system in the rotating frame of a rotating crystal. The normal modes are then found by harmonic expansion around the equilibrium state. Because the Penning trap creates a rotating crystal, the normal modes must be solved in the rotating frame, and the longitudinal (planar) modes require the solution of a quadratic eigenvalue problem because they have velocity-dependent forces. The phonon Hamiltonian is then derived by quantizing the normal modes.
In Sec. II, we also discuss quantum simulation based on spin-dependent forces arising from laser-ion dipole interactions. The effective Ising models
mediated by the axial phonon modes and the planar phonon modes are also described.
In Sec. III, we show the numerical results for phonon frequencies and for spin-spin interactions for
parameters relevant for current experiments. In Sec. IV, we provide our conclusions.

\begin{figure}[http!]
\centering
\includegraphics[scale=0.4]{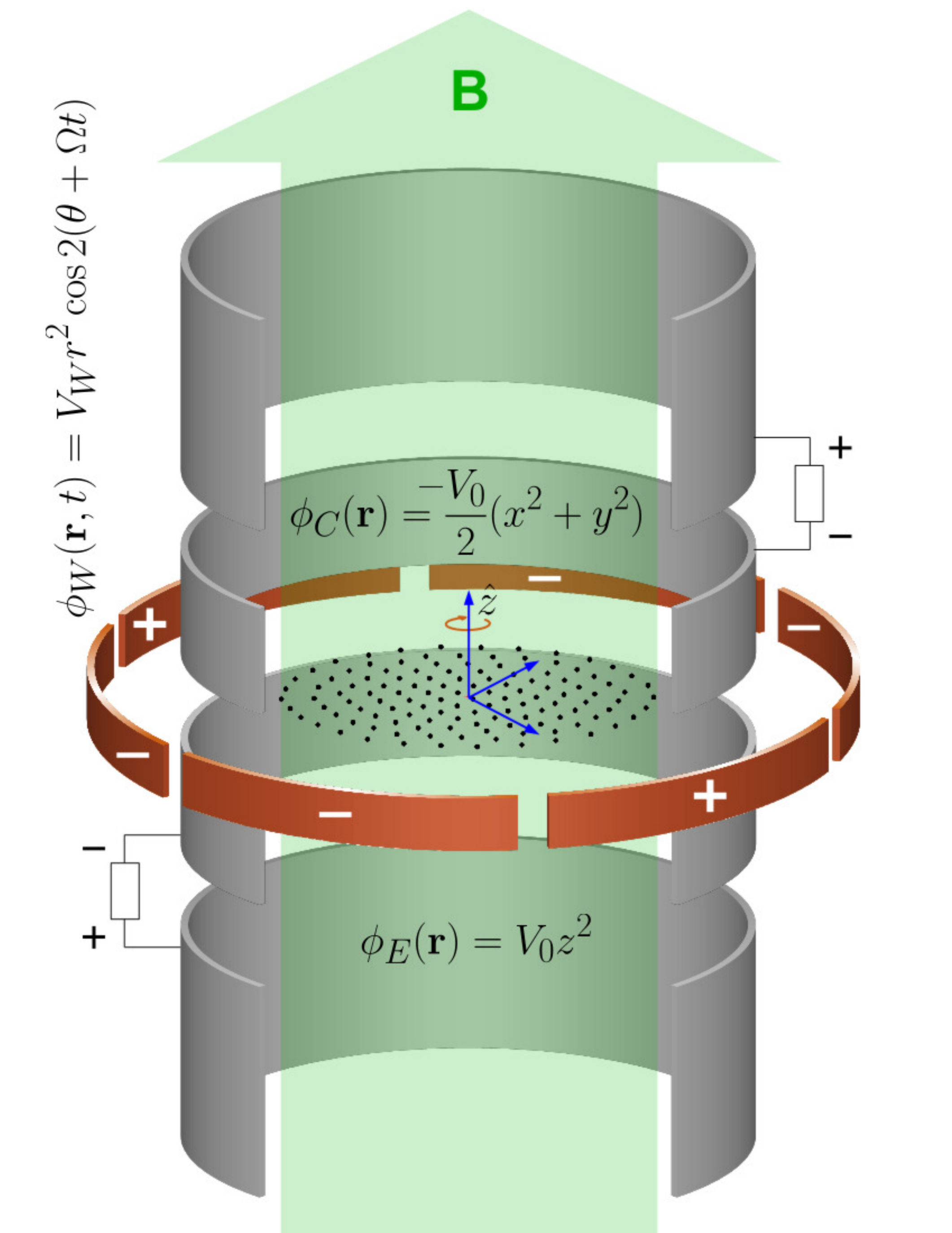}\\
\caption{Electrostatic potentials  provided by electrodes in a Penning trap have contributions from both the end-cap electrodes [with the trapping potential $\phi_{E}({\bf r})=V_{0}z^{2}$ that pushes the atoms towards $z=0$] and the cylindrical electrodes [with the radial quadratic potential $\phi_{C}({\bf r})=-\frac{V_{0}}{2}(x^2+y^2)$, which tends to push the ions out radially]. In addition, the repulsive Coulomb potential between the ions tends to destabilize the system in the trap.
A static magnetic field ${\bf B}=B_{z}{\hat z}$($B_{z}>0$) is thus applied along the axial direction to provide the radial confinement of the ions.
To lock the rotational angular frequency of the ions at a specific rotational speed, a time-dependent clockwise quadrupole potential $\phi_{W}(t)=V_{W} r^{2} \cos2(\theta +\Omega t)$ ($\Omega > 0$) is applied to the ions through ring electrodes so that the steady state of the ions (with a rigid body rotational speed $\Omega$) can be phase locked. Note that the rotating quadrupole potential is well implemented by
 ring electrodes with just six sectors (as employed in the NIST Penning traps~\cite{John,John2,rotatingwall}).}
  \label{fig:FIG1}
\end{figure}

\section{Theoretical formulation}
We consider hundreds of spins, realized via
the two hyperfine states $|{}^{2}\textrm{S}_{1/2},m_{J}=1/2\rangle$,
$|{}^{2}\textrm{S}_{1/2},m_{J}=-1/2\rangle$ of ${}^{9}\textrm{Be}^{+}$ ions, localized in a two-dimensional plane. The setup of cold ions in Penning traps is briefly illustrated in Fig.~1. Further details can be found in the published literature~\cite{John,John2}.
 In current experiments, there also is a small amount of impurities such as $\textrm{BeH}^{+}$ which forms via collision of the Beryllium ions with hydrogen molecules that are in the background gas. Hence a real system will have different mass particles (which will tend to separate from one another as the centrifugal force pushes the heavier particles to the outer regions of the crystal). In this paper, we discuss the clean limit only, and ignore such defects. The effect of defects will be studied elsewhere.

Given the above considerations, we form the ion Lagrangian in the laboratory frame of reference as
\begin{equation}
\label{eq:Lagrangian_Lab}
L=\sum_{j=1}^{N}\left[\frac{1}{2}m|\dot{{\bf r}}_{j}|^{2}-e\phi_{j}+e{\bf A}_{j}\cdot \dot{{\bf r}}_{j}\right],
\end{equation}
in which $N$ is the total number of ions, $e$ is the positive unit charge of and $m$ is the mass of a
 ${}^{9}\textrm{Be}^{+}$ ion, ${\bf r}_{j}=(\rho_{j}, \theta_{j}, z_{j})$ is the ion spatial configuration in cylindrical coordinates, $\phi_{j}(t)$ is the total  scalar potential acting on the $j$th ion and ${\bf A}_{j}=\frac{1}{2}({\bf B}\times {\bf r}_{j})$
is the vector  potential in the symmetric gauge for the axial magnetic field ${\bf B}=B_{z}{\hat z}~(B_{z} > 0)$.
The potential $\phi_{j}$, includes
the harmonic trapping from the electrodes, the rotating wall potential, and
 the Coulomb interaction between the ions. It satisfies
\begin{equation}
\phi_{j}(t)=V_{0}\left[z_{j}^{2}-\frac{1}{2}r_{j}^{2}\right] +V_{W} r_{j}^{2} \cos[2(\theta_{i} +\Omega t)]+\frac{k_{e}e^{2}}{2}\sum_{k\neq j}\frac{1}{r_{kj}},
\end{equation}
in which $V_0$ is the amplitude of the static potential from the Penning trap electrodes, $V_{W}$ is the amplitude of the rotating wall potential, $\Omega > 0 $ is the rotating wall angular frequency ${\bf \Omega}=\Omega {\hat z}$, $k_{e}$ is the Coulomb force constant, and $ r_{kj}=|{\bf r}_k-{\bf r}_j|$ is the inter-particle distance between the $k$th and $j$th ions.
With the application of the quadrupole rotating wall potential $\phi_{W}(t)$, the ion potential $\phi_{j}(t)$ is time dependent in the laboratory frame of reference.
However, in the co-rotating frame with the angular speed $\Omega$, the corresponding effective potential becomes time independent and the problem can be understood as an equilibrium problem~\cite{dubin_oneil}.
The geometric configuration of the ions under rotation in the steady state is simplified
by finding the (static) equilibrium ion configuration in the rotating frame.
In general, the coordinate transformation between the rotating frame and the laboratory frame is described by the $O(2)$ rotational operator $R_{z}(\theta(t))$ with the angle of rotation $\theta(t)$:
\begin{equation}
\label{eq:transformation}
\left(
  \begin{array}{c}
    x_{j}(t) \\
    y_{j}(t) \\
  \end{array}
\right) =R_{z}(\theta(t))
         \left(
           \begin{array}{c}
             x_{j}^{R}(t) \\
             y_{j}^{R}(t) \\
           \end{array}
         \right)
\end{equation}
where the superscript $R$ denotes the rotating frame of reference.
The operator $R_{z}(\theta(t))$ satisfies
\begin{equation}
R_{z}(-\Omega t)=\left(
           \begin{array}{cc}
             \cos(\Omega t) & \sin(\Omega t) \\
             -\sin(\Omega t) & \cos(\Omega t) \\
           \end{array}
         \right),
\end{equation}
with $\theta(t)=-\Omega t$.
Accordingly, in the rotating frame, the rotating wall potential $\phi_{W}^{R}$ does not depend on time and is given by
$\phi_{W}^{R}=\sum_{j}V_{W}({x_{j}^{R}}^{2}-{y_{j}^{R}}^{2})$
by direct substitution. As a consequence, the effective  potential energy in the rotating frame is then given by the expression
\begin{widetext}
\begin{equation}
e\phi_{j}^{R}=
eV_{0}{z_{j}^{R}}^{2}+\frac{1}{2}(eB_{z}\Omega-m\Omega^{2}-eV_{0}
){r_{j}^{R}}^{2}
+eV_{W}({x_{j}^{R}}^{2}-{y_{j}^{R}}^{2})
+\frac{ke^{2}}{2}\sum_{k\neq j}\frac{1}{ r_{jk}^{R}}
\label{eq: phi_r}
\end{equation}
\end{widetext}
which now includes the potential terms from the centrifugal and Lorentz forces as well.

Ions trapped in a Penning trap do not always crystallize into a two-dimensional plane. The following approximate confinement criterion has to be satisfied to guarantee it is energetically favorable for ions to stay in the plane with $z=0$:
\begin{equation}
\beta_{1}=\frac{2eV_{0}}{(eB_{z}\Omega-m\Omega^{2}-eV_{0})}\gg 1.
\end{equation}
In addition, when the rotating wall potential vanishes ($V_{W}\rightarrow 0$), the system is confined in the plane only if the radial potential is attractive. In this case, the confinement criterion for zero rotating wall potential is given by
\begin{equation}
\beta_{2}=eB_{z}\Omega-m\Omega^{2}-eV_{0}>0.
\end{equation}
For nonzero rotating wall potentials, the ion saddle potential caused by the rotating wall potential $V_{W}$ breaks rotational symmetry and leads to a confinement of ions only when the
coefficients of the trapping potential along the weak trapping axis is positive.
In other words, the following criterion has to be satisfied to have
a confined equilibrium state for all ions:
\begin{equation}
  \beta_{3}= \frac{1}{2}(eB_{z}\Omega-m\Omega^{2}-eV_{0})-e|V_{W}|
   > 0.
\end{equation}
 The above heuristic criteria narrow down the relevant parameter regime in which a stable planar ion crystal formation is feasible for quantum simulation. Of course, the accurate quantitative prediction of the stability of the ion crystals must also include the Coulomb interaction between the ions. The precise criterion can only be found by solving for the equilibrium positions and showing that all normal modes have real frequencies, so that the equilibrium is globally stable~\cite{dubin1}.

Based on the invariance of the Lagrangian,
we derive the Lagrangian  in the rotating coordinates from Eq.~(\ref{eq:Lagrangian_Lab}) as
\begin{equation}
\label{eq:Lagrangian}
L^{R}=\sum_{j=1}^{N}\left[\frac{1}{2} m|\dot{{\bf r}}_{j}^{R}|^{2}-\frac{eB^{eff}[\Omega]}{2}({\dot x}_{    j}^{R}y_{j}^{R}-{\dot y}_{j}^{R}x_{j}^{R})-e\phi_{j}^{R}\right]
\end{equation}
in which the second term will produce the Lorentz force term with the effective magnetic field ${B}^{eff}[\Omega]\hat z$ for the $j$th ion
in the rotating coordinates with the magnitude $B^{eff}[\Omega]=B_{z}-2\Omega m/e$ which depends on the rotational angular frequency $\Omega>0$. The modification of the magnetic field is due to the Coriolis
force for the moving ions in the rotating frame. There is no Coriolis force
present for ions when the ions are in equilibrium in the rotating frame but the Coriolis force can have effects on the oscillating normal mode motion away from equilibrium.
The centrifugal force originates from
the term $-\frac{1}{2}m\Omega^2{r_{j}^R}^2$ in the effective potential energy $e\phi_{j}^{R}$ in Eq.~(\ref{eq: phi_r}).

\subsection{Stable configurations and normal modes in the rotating frame}
It has been well established experimentally that ions in Penning traps reach different static equilibrium states in the rotating frame of reference (steady state in the laboratory frame of reference) for different values of the adjustable rotating angular frequency given by a rotating wall potential $\phi_{W}$~\cite{John3}. Therefore, we can
find the stable spatial configuration of the ions by determining the local minima of the effective potential energy in the rotating frame of reference. In general, such a minimization problem is
difficult to solve for the absolute minimum in two or higher dimensions, because many different configurations can be competitive for the lowest energy state and there can be large potential barriers between them.  To find an experimentally viable solution, we are guided by the experiments themselves, which typically observe the ions arranged in an approximate triangular lattice in a single plane.

Our strategy is to construct the trial configurations based on a ``closed shell'' construction analogous to finding stable electronic configurations in an atom as detailed in the numerical discussions.
Even though we cannot guarantee that the stable configuration we find is the global minimum, the NIST experimental systems seem to find the same local minimum which validates our approach. In further support of the strategy, we have successfully predicted the phonon mode spectra and spin-spin interaction observed in the NIST experiments~\cite{John,Brian}.

In the following, we discuss the collective phonon excitations about the previously determined equilibrium configuration.
Based on the equilibrium configuration, normal mode dynamics of ions near the equilibrium structure
can be fully captured by the full system Lagrangian $L$
expanded around this minimal ion spatial configuration to quadratic orders by a Taylor series. We use the following notation for the ion coordinates ${\bf r}_{j}(t)={\bf R}_{j}^{0}+{\bf \delta R}_{j}(t)$
and for the ion velocities ${\dot{\bf r}}_{j}(t)={\dot{\bf \delta R}}_{j}(t)$ in the rotating frame.
Since the equilibrium configuration is the local minimum of the classical action of the system $S = \int dt L$, the first order terms in ${\bf \delta R}_{j}(t)$ and ${\dot{\bf \delta R}}_{j}(t)$
do not contribute to the expansion by construction. Therefore, the system Lagrangian can be expanded to quadratic order as
\begin{widetext}
\begin{equation}
L=L_{0}+\frac{1}{2}\sum_{j,k=1}^{N}\left [ ({\bf \delta R}_{j}\cdot \frac{\partial}{\partial {\bf R}_{j}})({\bf \delta R}_{k}\cdot \frac{\partial}{\partial {\bf R}_{k}})+(\dot{{\bf \delta R}}_{j}\cdot \frac{\partial}{\partial \dot{{\bf R}}_{j}})(\dot{{\bf \delta R}}_{k}\cdot \frac{\partial}{\partial \dot{{\bf R}}_{k}})+2({{\bf \delta R}}_{j}\cdot \frac{\partial}{\partial {{\bf R}}_{j}})(\dot{{\bf \delta R}}_{k}\cdot \frac{\partial}{\partial \dot{{\bf R}}_{k}})\right ] L{\Big{|}}_{0}
\end{equation}
\end{widetext}
 where the first term $L_{0}$ is the Lagrangian from the equilibrium state and the second term is the Lagrangian $L_{ph}$ due to fluctuations away from the equilibrium state.
One can isolate the planar  motions of the ions in the $xy$ plane from the axial motions along the $z$ axis
based on the consideration that the Lorentz force due to the external magnetic fields acts only in the crystal plane and the potentials are separable in cylindrical coordinates, hence
there is no harmonic coupling between planar and axial degrees of freedom (anharmonic effects can couple normal modes together but are ignored here). Therefore, we can study the axial and planar modes
independently.

The Lagrangian which governs the axial motions of the ions is described as
\begin{equation}
L_{ph}^{A}= \frac{1}{2}\sum_{j=1}^{N}m{\delta\dot{R}_{j}^{z}}{\delta\dot{R}_{j}^{z}}
-\frac{1}{2}\sum_{j,k=1}^{N}{K}_{jk}^{zz}\delta R_{j}^{z}\delta R_{k}^{z},
\end{equation}
in which the symmetric stiffness matrix ${\bf K}^{zz}$ is a real Hermitian matrix $K_{jk}^{zz}=K_{kj}^{zz}$ and the matrix elements $K_{jk}^{zz}$ are given by
\begin{equation}
K_{jk}^{zz}=\left\{
\begin{array}{lll}
    \displaystyle {2eV_{0}-\sum_{l=1}^{N}\frac{k_{e}e^{2}}{{R_{l j}^{0}}^{3}}} &\ \ \ \ j=k, l \neq j\\
    \displaystyle {\frac{k_ee^2}{{R_{jk}^{0}}^{3}}} &\ \ \ \ j \neq k,
\end{array}\right\}
\end{equation}
where $R_{jk}^{0}=|{\bf R}_j^0-{\bf R}_k^0|$ is the distance between two ions in equilibrium (in the rotating frame).
Similarly, we derive the Lagrangian for the planar normal modes
\begin{equation}
\label{eq:Lagrangian_planar}
\begin{array}{ll}
\displaystyle{L_{ph}^{P} = \frac{1}{2}\sum_{j=1}^{N}m|\dot{\bf \delta R}_{j}|^{2}-\frac{1}{2}
\sum_{j,k=1}^{N}\sum_{\alpha\beta\in (x,y)} K_{jk}^{\alpha\beta} \delta R_{j}^{\alpha}
\delta R_{k}^{\beta}} \\
\displaystyle{+\sum_{j=1}^{N}\frac{eB^{eff}[\Omega]}{2}
\left[\delta R_{j}^{x}
\delta{\dot R}_{j}^{y}-\delta R_{j}^{y}\delta{\dot R}_{j}^{x}\right]},
\end{array}
\end{equation}
in which $B^{eff}[\Omega]=B_{z}-2\Omega m/e$ is the effective magnetic field in the rotating frame with angular frequency $\Omega$.
The real Hermitian stiffness matrix ${\bf K}^{\alpha\beta}$ satisfies the relation $K_{jk}^{\alpha\beta}=K_{kj}^{\alpha\beta}=K_{jk}^{\beta\alpha}=K_{kj}^{\beta\alpha}$. We derive the stiffness matrix ${\bf K}^{\alpha\beta}$  for the planar modes as:
\[
K_{jk}^{xx}=\left\{
\begin{array}{llll}
 \displaystyle{m\Omega^2+e\Omega B^{eff}[\Omega]-eV_{0}+2eV_{W}}
 &\\
  \displaystyle{-k_{e}e^{2}\sum_{l=1}^{N}
   \frac{{R_{jl}^{0}}^{2}-3(x_{j}^{0}-x_{l}^{0})^{2}}{
    {R_{jl}^{0}}^{5}}} & \ \ \ \ j=k,l\neq j \\
       &   \\
   \displaystyle{k_{e}e^{2}\frac{{R_{jk}^{0}}^{2}-3(x_{j}^{0}-x_{k}^{0})^{2}}
  {{R_{jk}^{0}}^{5}}} & \ \ \ \ j \neq k,
\end{array}\right.
\]
\[
K_{jk}^{yy}=\left\{
\begin{array}{llll}
 \displaystyle{m\Omega^2+e\Omega B^{eff}[\Omega]-eV_{0}-2eV_{W}}
 &\\
  \displaystyle{-k_{e}e^{2}\sum_{l=1}^{N}
   \frac{{R_{jl}^{0}}^{2}-3(y_{j}^{0}-y_{l}^{0})^{2}}{
    {R_{jl}^{0}}^{5}}} & \ \ \ \ j=k,l\neq j \\
       &   \\
   \displaystyle{k_{e}e^{2}\frac{{R_{jk}^{0}}^{2}-3(y_{j}^{0}-y_{k}^{0})^{2}}
  {{R_{jk}^{0}}^{5}}} & \ \ \ \ j \neq k,
\end{array}\right.
\]

\begin{equation}
K_{jk}^{xy}=\left\{
\begin{array}{ll}
  \displaystyle{3k_{e}e^2\sum_{l=1}^{N}
  \frac{(x_{j}^{0}-x_{l}^{0})(y_{j}^{0}-y_{l}^{0})}{{R_{jl}^{0}}^{5}}} \ \ \  &  \ \ \ \ j=k,l\neq j \\
  & \\
\displaystyle{-3k_{e}e^2\frac{(x_{j}^{0}-x_{k}^{0})(y_{j}^{0}-y_{k}^{0})}
{{R_{jk}^{0}}^{5}}}& \ \ \ \ j \neq k
\end{array}\right.
\label{eq: planark}
\end{equation}
where the equilibrium configuration is represented by the ion coordinates ${\bf R}_{j}^{0}=(x_{j}^{0}, y_{j}^{0}, z_{j}^{0})$ for $j=1,2,...,N$.
Notice that the off-diagonal matrix elements ${\bf K}^{xy}$ have nonzero values,
indicating the collective nature of the planar motional degrees of freedom, which couple the motion in the two coordinate directions together.

\subsection{Axial phonon modes}
Phonon modes are the quantized modes corresponding to the classical normal modes discussed earlier.
In this section, we solve the classical normal mode problem and then find the quantized Hamiltonian of the axial phonons, which is most relevant for
the description of conventional quantum simulation in a Penning trap.

By minimizing the action $\delta \int dt L_{ph}^{A}=0$, the classical equation of motion for the ion displacement along the axial ($+z$) axis can be written as
\begin{equation}
\label{eq:TRANSVERSE}
m\delta\ddot{R}_{j}^{z}+\sum_{k=1}^{N}K_{jk}^{zz}\delta R_{k}^{z}=0,\  j=1,2, \ldots N .
\end{equation}
The axial normal modes are found by direct substitution into
Eq.~(\ref{eq:TRANSVERSE}) of the eigenvector solution $\delta R_j^z(t)= b_{j}^{z\nu}\cos [\omega_{z\nu}(t-t_{0})]$ after initial time $t_{0}$ where the eigenvectors $b_{j}^{z\nu}$ are $N$-tuples of real numbers with unit norm.
The following eigenvalue equation must therefore be satisfied
\begin{equation}
\sum_{k=1}^{N}\left[m\omega_{z\nu}^{2}
\delta_{jk}-K_{jk}^{zz}\right]b_{k}^{z\nu}=0,\ j,\nu=1,2, \ldots N
\end{equation}
where the $N$ axial eigenvectors are the eigenvectors of the stiffness matrix ${\bf K}^{zz}$ (whose
eigenvalues are $\lambda_{z\nu}$)
but with different eigenvalues $\omega_{z\nu}=\sqrt{\lambda_{z\nu}/m}$ (since all eigenvalues $\lambda_{z\nu}$ are real, the frequencies are either real or pure imaginary). Note that this eigenvalue problem is the simplest quadratic eigenvalue problem, but because one can immediately solve it by this mapping onto a linear Hermitian eigenvalue problem, we need not discuss this point further here.  For the planar phonons, the analysis is more complicated, as shown below.

The planar crystal system is stable against a one-to-two plane transition when all the eigenfrequencies $\omega_{z\nu}$ are real, which is analogous to case of stability against the well-known zigzag transition for ions in linear Paul traps.
If the eigenfrequencies $\omega_{z\nu}$ are imaginary, then the two-dimensional planar equilibrium that
was found is an unstable equilibrium, and will not form in experiment.
A detailed numerical analysis of this instability can tell us the parameter regimes where a single sheet of ion crystals is energetically stable~\cite{dubin1}.

The procedure for the quantization of axial normal modes is the standard procedure.
One identifies the generalized coordinates $Q_{\nu}$ and momentum $P_{\nu}$ for each
phonon mode $\nu$ as canonically conjugate variables,
which satisfy the relation given by the Poisson bracket  $\{ Q_{\nu},P_{\nu'}\}=\delta_{\nu\nu'}$.
The quantization of phonon modes can be implemented by promoting the relation $\{Q_{\nu},P_{\nu'}\}=\delta_{\nu\nu'}$ to the commutation relations $[\hat Q_{\nu},\hat P_{\nu'}]=i\hbar\delta_{\nu\nu'}$ for operators.
To identify the canonically conjugate variables for the axial motional degrees of freedom,
the Lagrangian for the axial phonon modes
can be recast into the  form in Eq.~(\ref{eq: axial_qm_ham}) (as long as the system is in the parameter regime in which the planar crystal configuration is stable with Im$[\omega_{z\nu}]=0$ for all modes $\nu$). We let $\delta R_{j}^{z}(t)=\sum_{\nu}\xi_{\nu}(t)b_{j}^{z\nu}$ be the displacements, where the quantum dynamics due to the phonon mode $\nu$ is given by the real generalized coordinate $\xi_{\nu}(t)$.  The Lagrangian becomes
 \begin{equation}
   L_{ph}^{A} = \frac{1}{2}\sum_{\nu=1}^{N} m\left[{\dot\xi}_{\nu}^2
 -\omega_{z\nu}^{2}\xi_{\nu}^2\right].
\label{eq: axial_qm_ham}
 \end{equation}

The corresponding generalized momentum for the phonon mode $\nu$  is then given by the expression
\begin{equation}
P_{\nu}^{A}=\frac{\partial L_{ph}^{L}}{\partial {{\dot \xi}_{\nu}}}=m{{\dot{\xi}}_{\nu}}.
\end{equation}
As a consequence, the Hamiltonian for the axial phonon modes $H_{ph}^{A}$
is represented by the summation over $N$ independent harmonic modes
\begin{equation}
H_{ph}^{A}=\sum_{\nu=1}^{N}\left[\frac{{{ P}_{\nu}^{A ^2}}}{2m}+\frac{1}{2}m\omega_{z\nu}^{2}
 {{\xi}_{\nu}^2}\right].
\end{equation}
Therefore, the second quantized form of the Hamiltonian ${\hat H}_{ph}^{A}$ for the axial phonon modes simply becomes
\begin{equation}
{\hat H}_{ph}^{A}=\sum_{\nu=1}^{N}\hbar\omega_{z\nu}(\hat{n}_{z\nu}+\frac{1}{2}),
\label{eq: phonon_axial_ham}
\end{equation}
in which $\hat{n}_{z\nu}=\hat a_{z\nu}^{\dagger}\hat a_{z\nu}$ is the number operator for the phonon mode $\nu$ and the phonon creation and annihilation operators are related to the generalized coordinates via
\begin{eqnarray}
   a_{z\nu} &=&\sqrt{\frac{m\omega_{z\nu}}{2\hbar}}\left(\xi_{\nu}+\frac{i}{m\omega_{z\nu}}{ P}_{\nu}^{ A}\right), \nonumber \\
    a_{z\nu}^{\dagger} &=& \sqrt{\frac{m\omega_{z\nu}}{2\hbar}}\left(\xi_{\nu}-\frac{i}{m\omega_{z\nu}}{ P}_{\nu}^{ A}\right)
\end{eqnarray}

The phonon creation and annihilation operators are related to the quantum fluctuations of ions along the axial direction via
\begin{equation}
\delta {\hat R}_{j}^{z}=\sum_{\nu}  b_{j}^{z \nu}\sqrt{\frac{\hbar}{2m\omega_{z\nu}}}[\hat a_{z \nu}+\hat a^{\dag}_{z \nu}].
\end{equation}

It is obvious that the Hamiltonian ${\hat H}_{ph}^{A}$ is invariant
in the rotating frame and the laboratory frame
since each phonon mode oscillates along the rotation axis ${\hat z}$ and hence they are not influenced by the overall rotation of the coordinate system.

\subsection{Planar phonon modes}
So far, the planar phonon modes have not been utilized
for quantum simulation due to the complexity of the normal modes
as well as the complication introduced by the rotation of the ion crystals as observed in the laboratory frame.
We plan to give a detailed explanation of the nature of the planar phonon modes and hope this will stimulate further insights toward quantum simulation. Our procedure is based on previous work on Coulomb crystals in a magnetic field, but not in a rotating frame~\cite{japanese,russian_old,chen_thesis,baiko}.  Let us expand the Lagrangian $L_{ph}^{P}$ in terms of the $2N$ eigenmodes
of the hermitian stiffness matrix ${\bf K}$ in Eq.~(\ref{eq: planark}), which are labeled by the index $\nu$ with $1\le\nu\le 2N$ (note that these eigenvectors are {\it not} the normal mode eigenvectors for the planar motion, they are a convenient basis to use for the normal mode equations of the planar motion, except for the one case noted below). These eigenvectors $b_{k}^{\alpha \nu}$  satisfy
\begin{equation}
\sum_{k \beta}K_{jk}^{\alpha\beta}b_{k}^{\beta \nu}=m{\omega_{0}^{\nu}}^2 b_{j}^{\alpha \nu},
\end{equation}
  where the spatial indices $j=1,\cdots N, k=1,\cdots N$ are the site indices for each ion and $\alpha, \beta=x$ or $y$, denote the directions of the spatial displacements of the phonons in the plane, $m{\omega_{0}^{\nu}}^{2}$ are the real eigenvalues of the real-symmetric matrix ${\bf K}$ for the stable ion configuration. These eigenvectors do become the eigenvectors for the planar modes only in the case when $B_{eff}=0$ (which occurs when the rotating frequency is equal to half the cyclotron frequency), where the eigenvalue problem simplifies to the traditional form for a phonon. This fact has been used for an analysis of the phonons in a Penning trap~\cite{Jake}. The basis vectors $b_{k}^{\alpha \nu}$ can be chosen to be real and they are also orthonormal $\sum_{i\alpha}b_i^{\alpha\nu}b_i^{\alpha\nu'}=\delta_{\nu\nu'}$. Therefore, the component of the displacement of the $j$th ion in the $\alpha$th direction, $\delta R_{j}^{\alpha}(t)$, can be expanded in this basis as
 \begin{equation}
 \delta R_{j}^{\alpha}(t)=\sum_{\nu=1}^{2N}\zeta_{\nu}(t)b_{j}^{\alpha \nu},~ j=1,2,\cdots N.
 \end{equation}
where $\zeta_{\nu}(t)$ is the dynamic variable.
The Lagrangian $L_{ph}^{P}$ for the planar normal modes in this basis can be rewritten following Eq.~(\ref{eq:Lagrangian_planar}) as
\begin{equation}
L_{ph}^{P} = \frac{1}{2}m \left[\dot{\zeta}_{\nu}^{2}
-{\omega_{0}^{\nu}}^2{\zeta_{\nu}}^2\right]-\frac{1}{2}\zeta_{\nu}
\dot{\zeta}_{\nu'}T_{\nu\nu'},
\end{equation}
where the Einstein summation convention is used for repeated indices and
the matrix elements $T_{\nu\nu'}$ are related to the antisymmetric matrix $T_{jk}^{\alpha\beta}$  by the unitary transformation $T_{\nu\nu'}=b_{j}^{\alpha\nu}T_{jk}^{\alpha\beta}b_{k}^{\beta\nu'}$ in which $T_{jk}^{\alpha\beta}=-\frac{e}{2}B_{eff}[\Omega] \delta_{jk}\epsilon_{\alpha\beta}=-T_{kj}^{\beta\alpha}$ (where $\epsilon_{xy}=-\epsilon_{yx}=1$ and all others vanish)  due to the velocity dependent term in Eq.~(\ref{eq:Lagrangian_planar}).
Based on the Lagrangian $L_{ph}^{P}$, we can identify that the canonical momentum $P^{\nu}$ for mode $\nu$ is related to the mechanical momentum
$\Pi^{\nu}=m\dot{\zeta}_{\nu}$ by
\begin{equation}
P^{\nu}=\frac{\partial{L_{ph}^{P}}}{{\partial \dot{\zeta}_{\nu}}}=\Pi^{\nu}-\frac{1}{2}\zeta_{\nu'}b_{j}^{\alpha\nu'}
T_{jk}^{\alpha\beta}b_{k}^{\beta\nu}=\Pi^{\nu}-\frac{1}{2}\zeta_{\nu'}T_{\nu'\nu}
\end{equation}
Finally, we arrive at the Hamiltonian  for the planar modes as
\begin{equation}
H_{ph}^{P}=P^{\nu}\dot{\zeta}_{\nu}-L_{ph}^{P}=\frac{1}{2m}{\Pi^{\nu}}^2+\frac{1}{2}m{\omega_{0}^{\nu}}^{2}
{\zeta_{\nu}}^2.
\end{equation}
Notice that the Hamiltonian $H_{ph}^{P}$ is not diagonal in the canonical conjugate variables $P^{\nu}$ and $\zeta_{\nu}$ but is diagonal in the variables  $\Pi^{\nu}$ and $\zeta_{\nu}$. We need to perform a canonical transformation in order to diagonalize the Hamiltonian $H_{ph}^{P}$ in the new canonical conjugate variables and we then can proceed with the standard quantization procedure (as in the axial phonon case).

Our strategy is to first quantize the Hamiltonian with respect to the canonical variables and then choose a nonstandard linear combination of coordinate and momentum to create the appropriate raising and lowering operators (this strategy was described by Baiko~\cite{baiko}). Hence we use the following fundamental commutation relations
\begin{equation}
[\zeta_{\nu},\zeta_{\nu'}]=0, \ [P^{\nu},P^{\nu'}]=0,\  [\zeta_{\nu}, P^{\nu'}]=i\hbar\delta_{\nu\nu'},
\end{equation}
which lead to the following commutation relations for the mechanical momenta
\begin{eqnarray}
  [\Pi^{\nu},\Pi^{\nu'}] &=&
[P^{\nu},\frac{1}{2}\zeta_{\bar \nu}b_{j}^{\alpha\bar \nu}T_{jk}^{\alpha\beta}b_{k}^{\beta\nu'}]+
[\frac{1}{2}\zeta_{\bar \nu}b_{j}^{\alpha\bar \nu}T_{jk}^{\alpha\beta}b_{k}^{\beta\nu},P^{\nu'}] \nonumber \\
   &=& -i\hbar T_{\nu\nu'}\neq 0,
\end{eqnarray}
in which the antisymmetric property $T_{\nu\nu'}
=b_{j}^{\alpha\nu}T_{jk}^{\alpha\beta}b_{k}^{\beta\nu'}
=-b_{k}^{\alpha\nu'}T_{kj}^{\beta\alpha}b_{j}^{\beta\nu}=
-T_{\nu'\nu}$ is used.
To diagonize the Hamiltonian in the conventional second quantized form
$H_{ph}^{P}=\sum_{\lambda=1}^{2N}
\hbar\omega_{\lambda}(a_{\lambda}^{\dagger}a_{\lambda}+\frac{1}{2})$
with $2N$ collective mode frequencies $\omega_{\lambda}>0$ for the stable ion configurations, we look for the phonon creation operator $a_{\lambda}^{\dagger}$ in the presence of the magnetic field
as the unconventional superposition of the operators $\Pi^{\nu}$ and $\zeta_{\nu}$
\begin{equation}
a_{\lambda}^{\dagger}=\alpha_{\lambda}^{\nu}\Pi^{\nu}+\beta_{\lambda}^{\nu}
\zeta_{\nu}, \quad a_{\lambda}=\alpha_{\lambda}^{\nu*}\Pi^{\nu}+\beta_{\lambda}^{\nu*}
\zeta_{\nu},
\label{eq:collective}
\end{equation}
with $\alpha$ and $\beta$ (possibly complex) numbers.
The operators $a_{\lambda}$ and $a_{\lambda}^{\dagger}$ must satisfy the commutation relation $[a_{\lambda},a_{\lambda'}^{\dagger}]=\delta_{\lambda\lambda'}$
As a consequence, the Hamiltonian $H_{ph}^{P}$ is required to satisfy the commutation relation
\begin{equation}
[H_{ph}^{P},a_{\lambda}^{\dagger}]=\hbar\omega_{\lambda}a_{\lambda}^{\dagger}
\label{eq:Ph_Hamiltonian}
\end{equation}
with $\omega_\lambda$ the normal mode frequency for the planar phonon.
By direct substitution of Eq.~(\ref{eq:collective}) into Eq.~(\ref{eq:Ph_Hamiltonian}), the coefficients $\alpha_{\lambda}^{\nu}$
and $\beta_{\lambda}^{\nu}$ must satisfy the following coupled eigenvalue equations
\begin{eqnarray}
\label{eq:eig_alpha}
  \omega_{\lambda}\alpha_{\lambda}^{\nu} &=& -\frac{i}{m}\beta_{\lambda}^{\nu}-\frac{i}{m}
  T_{\nu\nu'}\alpha_{\lambda}^{\nu'} \\
\label{eq:eig_beta}
  \omega_{\lambda}\beta_{\lambda}^{\nu} &=& i m {\omega_{0}^{\nu}}^2 \alpha_{\lambda}^{\nu}.
\end{eqnarray}
Eq.~({\ref{eq:eig_beta}}) can be solved for $\beta$ in terms of $\alpha$ via $\beta_{\lambda}^{\nu}=i m \frac{{\omega_{0}^{\nu}}^2}{\omega_{\lambda}}\alpha_{\lambda}^{\nu}$. Substituting into Eq.~(\ref{eq:eig_alpha}), we
 find the $\alpha$ coefficients satisfy a quadratic eigenvalue problem (QEP)
\begin{equation}
\Big(m\omega_{\lambda}^2\delta_{\nu\nu'}+i\omega_{\lambda}T_{\nu\nu'}
-m{\omega_{0}^{\nu}}^2\delta_{\nu\nu'}\Big)\alpha_{\lambda}^{\nu'}=0
\label{eq:equation_of_motion_planar}
\end{equation}
for the eigenvalue $\omega_\lambda$ which is chosen to be a nonnegative frequency. Since QEP are not so common in physics, we discuss how to solve them in the appendix (see also Ref.~\onlinecite{QEP}).  In the simple form in Eq.~(\ref{eq:equation_of_motion_planar}), we can map the QEP onto a conventional linear hermitian eigenvalue problem in twice as many dimensions, which then allows us to use orthonogonality of eigenvectors and completeness to derive a number of nontrivial identities of the $\alpha$ eigenfunctions. Details can be found in the appendix. For example, if we consider two different positive eigenvalues
 $\omega_{\lambda}$ and $\omega_{\lambda'}$, the corresponding eigenvectors satisfy the following orthogonality relation
\begin{equation}
\sum_{\nu=1}^{2N}\Big(\omega_{\lambda}\omega_{\lambda'}
+{\omega_{0}^{\nu}}^2\Big)\alpha_{\lambda}^{\nu*}\alpha_{\lambda'}^{\nu}=0, \ \lambda\neq\lambda',
\label{eq:ortho}
\end{equation}
which is derived in the appendix.
 The normalization of $\alpha_{\lambda}^{\nu}$ is fixed by the following commutation relation
\begin{eqnarray}
  [a_{\lambda},a_{\lambda'}^{\dagger}] &=& -i\hbar\alpha_{\lambda}^{\nu*}T_{\nu\nu'}\alpha_{\lambda'}^{\nu'}+
  \hbar m {\omega_{0}^{\nu}}^2 \Big( \frac{1}{\omega_{\lambda}}+\frac{1}{\omega_{\lambda'}}\Big)
  \alpha_{\lambda}^{\nu*}\alpha_{\lambda'}^{\nu}
   \nonumber \\
   &=& \hbar m \alpha_{\lambda}^{\nu*}
\alpha_{\lambda'}^{\nu}\Big[\frac{{\omega_{0}^{\nu}}^2}
{\omega_{\lambda}}+\omega_{\lambda'}\Big]=\delta_{\lambda\lambda'}
\label{eq:normalization}
\end{eqnarray}
which has been simplified by using  Eqs.~(\ref{eq:equation_of_motion_planar}) and (\ref{eq:ortho}).

We are interested in representing the operator $\zeta_{\nu}$ in terms
of the raising and lowering operators $a_{\lambda}$ and $a_{\lambda}^{\dagger}$
so we can relate the displacement  $\delta R_{j}^{\alpha}$
in the plane to the collective mode operators $a_{\lambda}$ and $ a_{\lambda}^{\dagger}$.
Using Eq.~(\ref{eq:collective}), $\zeta_{\nu}$ is related to $a_{\lambda}$ and $a_{\lambda}^{\dagger}$ via
\begin{eqnarray}
  \alpha_{\lambda}^{\nu*}a_{\lambda}^{\dagger}-
\alpha_{\lambda}^{\nu}a_{\lambda} &=& \Big(\alpha_{\lambda}^{\nu*}\beta_{\lambda}^{\nu}
-\alpha_{\lambda}^{\nu}\beta_{\lambda}^{\nu*}\Big)\zeta_{\nu}, \nonumber
 \\
   &=& 2 i \frac{m{\omega_{0}^{\nu}}^2}{\omega_{\lambda}}|\alpha_{\lambda}^{\nu}|^2
 \zeta_{\nu}, \\
 &=& \frac{i}{\hbar}\zeta_{\nu},
  \end{eqnarray}
in which the relation $\beta_{\lambda}^{\nu}=i m \omega_{0}^{\nu 2}\alpha_{\lambda}^{\nu}/{\omega_{\lambda}}$, Eq.~(\ref{eq: alpha_comp2}) and Eq.~(\ref{eq: alpha_comp3})
are used. Hence,
\begin{equation}
         \zeta_{\nu} = -i\hbar \sum_{\lambda}\Big(\alpha_{\lambda}^{\nu*}a_{\lambda}^{\dagger}
         -\alpha_{\lambda}^{\nu}a_{\lambda}\Big).
\end{equation}
This expression is crucial for the discussion of the planar modes in a quantum simulation.
The $\beta$ component of the ion displacement associated with phonon coherent state in Heisenberg picture can be evaluated as
\begin{eqnarray}
  \langle \delta R_{j}^{\lambda\beta} \rangle &=& 2\hbar\sum_{\lambda:\omega_\lambda>0}{\rm Re}\{i\alpha_{\lambda}^{\nu}b_{j}^{\alpha\nu}
\phi_{\lambda}e^{i\omega_{\lambda}t}\}\\
   &=&-2\hbar\sum_{\lambda:\omega_\lambda>0}
   |\phi_{\lambda}||\alpha_{j}^{\beta\nu}|
\sin(\omega_{\lambda}t+\delta_{\lambda})
\label{eq:planard}
\end{eqnarray}
  where $\phi_{\lambda}$ is the complex eigenvalue for the coherent state $|\phi\rangle$ ($a_{\lambda}|\phi\rangle=\phi_{\lambda}|\phi\rangle$) , $\delta_{\lambda}$ is the phase angle for mode $\lambda$, and $|\phi_{\lambda}|$ is the average phonon occupation for mode $\lambda$, $\alpha_{j}^{\beta\lambda}=\sum_\nu\alpha_\lambda^\nu b_j^{\beta \nu}$ is the projection of $\lambda$ state at site $j$ along the $\beta$ orientation.

\subsection{Effective spin Hamiltonian with axial phonon modes}

We discuss the generation of effective spin-spin couplings from an optical dipole force
within the context of the Penning trap experiment at NIST, which uses ions of Beryllium (modifications for other systems would be straightforward, but are omitted here).
Consider the hyperfine qubit states  $\mid \uparrow_{Z}\rangle$=$\mid  {}^{2}\textrm{S}_{1/2},m_{J}=1/2\rangle$ and $\mid \downarrow_{Z}\rangle$=$\mid{}^{2}\textrm{S}_{1/2},m_{J}=-1/2\rangle$ in a ${}^{9}\textrm{Be}^{+}$ ion. The qubit level splitting in the presence of a magnetic field of magnitude $B_{z}=4.46$ Tesla (due to Zeeman effects) is approximately $2\pi \times 124$ GHz. Even though the ${}^{9}\textrm{Be}^{+}$ nuclear spin is not a spin singlet,
the nuclear spin dynamics can be completely frozen out by optically pumping the nuclear spin to the single  nuclear spin state $m_{I} = +3/2$ throughout the duration of an experiment~\cite{John,John2,John3}, which is what we assume is done here.

A spin-dependent force along the $Z$ axis of the Bloch sphere is generated through a $\sigma^{Z}$ gate~\cite{PJ}, which is implemented by
a coupling of the qubit states with the excited states
($\mid\uparrow_{Z}\rangle \leftrightarrow \mid {}^{2}\textrm{P}_{3/2}, m_{J}\rangle$ and $\mid\downarrow_{Z}\rangle \leftrightarrow \mid {}^{2}\textrm{P}_{3/2}, m_{J}\rangle$)
 via two off-resonant laser beams with different angular frequencies $\omega_{U}, \omega_{L}$ and wavevectors $\bf {k_{U}},\bf {k_{L}}$ respectively.  The AC Stark shift due to the interference of two linear-polarized  beams
 for an ion labeled by $j$ generates the following spin-dependent force (after adjusting the polarization of the laser beams)~\cite{John}:
\begin{equation}
{\hat H}_{OD}=-\sum_{j=1}^{N}\frac{F_{O}}{\delta k_{z}}\cos(\delta k {\hat z}_{j}-\mu t)\sigma_j^Z,
\end{equation}
where the transverse wavenumber $|{\bf k_U-k_L}|=\delta k_{z} \approx 2k\sin(\theta_{R}/2)(|k_U|=k\approx|k_L|)$ is
determined by the magnitude of the wavevector difference between the two beams, $\theta_{R}/2$ is the angle between the beam with respect to the tangential orientation of the two-dimensional crystal plane, $\mu=\omega_{U}-\omega_{L}$ is the beatnote frequency given by the frequency difference of the two  beams, and ${\hat z}_{j}=\delta {\hat R}_{j}^{z}$ is the displacement of the transverse phonons along the axial direction $z$. The Pauli spin operator shows that the force is equal in magnitude, but opposite in direction for each of the two spin states of the Beryllium ion (hence it is known as a spin-dependent force).
In the Lamb-Dicke limit, we have $|\delta k_z\delta {\hat R}_{j}^{z}|\ll 1$, so the  interaction $H_{OD}$ can be further reduced to the following form
\begin{equation}
\label{eq:oda}
H_{OD}=-\sum_{j=1}^{N}F_{O}\delta {\hat R}_{j}^{z}\sin(\mu t)\sigma_{j}^{Z}
\end{equation}
after dropping a term with no $\delta {\hat R}_j^z$ dependence.
 In this limit, the spin-dependent force $-F_{O}\sin\mu t \sigma_{j}^{z}$ is spatially uniform.

The effective spin Hamiltonian and the time dependent evolution of the wave function from the Hamiltonian ${\hat H}_{OD}$ have been
rigorously studied when the system is cooled to low temperature to start from the phonon vacuum~\cite{spin-spin-interactions,Phonons} immediately before the onset of the quantum simulation. The evolution of the entangled spin-phonon states are captured by
the time evolution operator
\begin{equation}
U(t,t_{0})=\exp[-i\mathcal{H}^T_{ph}(t-t_0)/\hbar]\exp[-i {\hat W}_{I}(t)/\hbar] U_{spin}(t,t_0)
\end{equation}
with the operator ${\hat W}_I(t)$ determined by
${\hat W}_I(t)=\int_{t_0}^t dt^\prime {\hat V}_I(t^\prime)$ where the interaction
${\hat V}_{I}(t^\prime)$ is the optical dipole interaction term expressed in the Heisenberg
picture of the bare phonon states (also called the interaction picture)  via
\begin{equation}
{\hat V}_I(t)=\exp[i {\hat\mathcal{H}}_{ph}^{T} (t-t_0)/\hbar ]\mathcal{H}_{ODF}(t)\exp[-i {\hat \mathcal{H}}^{T}_{ph}(t-t_0)/\hbar],
\end{equation}
in which ${\hat \mathcal {H}}_{ph}^{T}$ is the (time-independent) bare planar phonon Hamiltonian given in Eq.~(\ref{eq: phonon_axial_ham}).
As a consequence, we find the effective spin Hamiltonian ${U}_{spin}(t,t_{0})$ is dictated by an
Ising spin Hamiltonian
\begin{equation}
\label{eq:Hspin}
{ U}_{spin}(t,t_{0})  =  \mathcal{T}_{t}\exp\left[-\frac{i}{\hbar}\int_{t_{0}}^{t}dt' \left(\sum_{j,j'=1}^N J_{jj'}(t')\sigma_{j}^{Z}\sigma_{j'}^{Z} \right)\right]
\end{equation}
with time-dependent exchange integrals.

These spin-spin exchange integrals are given by~\cite{spin-spin-interactions}
\begin{eqnarray}
  J_{jj'}(t) &=& \frac{F_{O}^2}{4m}\sum_{\nu=1}^{N}\frac{b_{j}^{z \nu}b_{j'}^{z \nu}}{\mu^2-\omega_{z \nu}^2} \nonumber \\
    & \times & [1+\cos{2\mu t}-\frac{2\mu}{\omega_{z\nu}}\sin{\omega_{z \nu} t}\sin{\mu t}].
  \label{eq:axialspin-spin}
\end{eqnarray}
The detailed phonon mode properties $\omega_{z\nu}$, $b_{j}^{z\nu}$, and $b_{j'}^{z\nu}$ determine the values of the different spin-spin interactions.
The sign of the effective spin-spin interaction $J_{jj'}(t)$ depends on the redness or blueness of the laser detuning $\mu$ with respect to each phonon mode and the range of the interaction can be tuned depending upon the magnitude of the laser detuning away from all the axial phonon modes.
In our numerical discussion, we will show these effects with direct numerical calculations.

Note that the analysis becomes more complicated if there is a transverse magnetic field present in addition to the spin-dependent optical dipole force.  This is needed for quantum simulations that employ adiabatic state creation. Nevertheless, we do not discuss the evolution of the system further here when there is a transverse magnetic field present. Details of the case of the Paul trap can be found in Ref.~\onlinecite{Phonons}.

\subsection{Spin-dependent interaction by planar modes}

Let us now discuss the situation where the momentum transfer from the two laser beams lies in the crystal plane. One complication of this set-up is due to the rotation of the crystal with respect to the laser beams in the laboratory frame of reference. Examining the planar-mode coupling is also important in order to estimate the errors for a quantum simulation mediated through the axial phonon modes  when the momentum difference $\hbar{(\bf k_U-k_L)}$ of the two Raman beams is not oriented precisely along the $z$ axis due to laser alignment errors. Nevertheless, one can also involve the planar modes on purpose for quantum simulation, which may be useful when the planar modes are far away from the axial modes in energy.
To simplify our discussion, let us choose the orientation of the planar spin-dependent forces due to the off-resonant laser beams to be along the $x$-axis in the laboratory frame.

Analogous to the earlier discussion for axial
phonon modes, the AC Stark shift along the $x$-axis for the planar phonon modes is associated with the effective momentum transfer  $\hbar \delta k_x \hat{x}$ due to the photons via
\begin{equation}
{\hat H}_{OD}=-\sum_{j=1}^{N}\frac{F_{O}}{\delta k_{x}}
\cos(\delta k_{x} {\hat x}_{j}(t)-\mu t)\sigma_j^Z.
\end{equation}
Notice that the ion coordinates ${\hat x}_{j}(t)=x_{j}^{0}(t)+\delta {\hat R}_{j}^{x}(t)$ in the laboratory frame are determined by
the rigid-body rotation of the ion crystals with the $x$ axis chosen along the orientation of the effective wavevector ${\bf k_U-k_L}=\delta k_x \hat{x}$ in the laboratory frame and fluctuations $\delta {\hat r}_{j}^{x}(t)$ in  the laboratory frame.
Hence, in the Lamb-Dicke limit $|\delta k_x\delta {\hat r}_{j}^{x}|\ll 1$, the optical dipole interaction ${\hat H}_{OD}$ is described by
\begin{equation}
\label{eq:HODP}
  {\hat H}_{OD} = \sum_{j=1}^{N}F_{O}\delta {\hat r}_{j}^{x}(t)\sin(\delta k_{x}x_{j}^{0}(t)-\mu t)\sigma_{j}^{Z}
\end{equation}
 where the phase $\delta k_{x}x_{j}^{0}(t)-\mu t$ for the planar modes
 is modulated by the rotation of the crystal
 for the ion coordinates in the laboratory frame given by $x_{j}^{0}(t)=R_{j}^{0}\cos(-\omega t+\phi_{j}^{0})$ with the static phase $\phi_{j}^{0}$ determined by the equilibrium configuration in the rotating frame with respect to the orientation of the spin-dependent force acting along the direction $\hat x$ at the initial time $t_{0}=0$.
Hence,
 the optical dipole interaction in Eq.~(\ref{eq:HODP}) can be expanded in harmonics of $\Omega$.
 We expect the effective spin-spin interaction due to the planar modes to be sensitive to the rotational angular frequencies and the laser detuning $\mu$. The value of the spin-spin interaction is sensitive to the structure of the different planar modes. Using the standard partial wave expansion~\cite{Partial}, we can expand the function $\sin(\delta k_{x}x_{j}^{0}(t)-\mu t)$ in  harmonics of $\Omega$ as
\begin{eqnarray}
   \sin(\delta k_{x}x_{j}^{0}(t)-\mu t)  &=& \sum_{l}f_{l}(t)j_{l}(\delta k_{x}R_{j}^{0}) \nonumber \\
    &\times &  P_{l}[\cos(-\Omega t +\phi_{j}^{0})]
 \end{eqnarray}
 where the function $f_l$ satisfies $f_{l}(t)=i^{l}(2l+1)\sin{\mu t}$ for an even integer $l$, and $f_{l}(t)=i^{l-1}(2l+1)\cos{\mu t}$ for an odd integer $l$. The function $j_{l}$ is the spherical Bessel function of the first kind, and
the function $P_{l}[\cos(-\Omega t +\phi_{j})]$ is the Legendre polynomial in harmonics of $\Omega$. Notice also that the fluctuations
$\delta {\hat r}_{j}^{x}$ are related to the fluctuations due to phonons in the rotating frame as $\delta {\hat r}_{j}^{x}=\delta {\hat R}_{j}^{x} \cos\Omega t +\delta {\hat R}_{j}^{y} \sin\Omega t $.
It is tempting to simplify the discussion further based on the belief
that one can tune the laser such that only one targeted planar mode is closest in energy.
Considering the narrow bandwidth of the planar modes, which is roughly an order of magnitude narrower than the bandwidth of the axial modes, the optical dipole interaction due to the planar phonon modes cannot usually be described by a single phonon mode even when the laser detuning is close to that mode. Therefore, we do not take this strategy for the planar modes.
In the following discussion, our derivation of the effective spin Hamiltonian is not restricted to any particular laser detuning or phonon band.

Following the procedure we mentioned earlier, the effective spin Hamiltonian due to the optical dipole interaction can be extracted by the commutation relation as
\begin{equation}
\label{eq:spin_hamiltonian}
{\hat H}_{spin}=\frac{i}{2\hbar}[{\hat W}_{I}(t),{\hat V}_{I}(t)],
\end{equation}
  in which the definition ${\hat V}_I(t)=\exp[i\hat{H}_{ph}^{P} t/\hbar ]\hat{H}_{OD}(t)\exp[-i \hat{H}_{ph}^{P} t/\hbar]$
and ${\hat W}_{I}(t)=\int_{0}^{t}dt'{\hat V}_{I}(t')$ are applied again.
After tedious algebra, we arrive at the following expression for the effective Ising spin Hamiltonian ${\hat H}_{spin}=\sum_{jj'}J_{jj'}(t)\sigma_{j}^{Z}\sigma_{j'}^{Z}$ with spin-spin interactions given by
\begin{eqnarray}
  J_{jj'}(t)&=& \hbar F_{O}^2\sum_{\nu \nu' l l' \beta \beta' \lambda}j_{l}(\delta k_x R_{j}^{0})j_{l'}(\delta k_x R_{j'}^{0})
  \nonumber \\
   &\times & b_{j}^{\beta \nu}b_{j'}^{\beta' \nu'}{\rm Im} \{\Big[g_{j'l'}^{\lambda' \beta}(t)+(1-\delta_{\beta\beta'})h_{j'l'}^{\lambda'\beta'}(t)\Big]
   \alpha_{\lambda}^{\nu'}\alpha_{\lambda}^{\nu*}\} \nonumber \\
   & \times& f_{l}(t)P_{l}[\cos(-\Omega t +\phi_{j}^{0})],
\end{eqnarray}
in which the indices
$\beta$, $\beta'$ run through $x,~y$ components of the normal modes
and the time dependent functions $h_{j,l}^{\lambda \beta}, g_{j,l}^{\lambda \beta}$ are defined by the following expressions
\begin{eqnarray}
  g_{j,l}^{\lambda x} &\equiv& \cos(\Omega t)\int_{0}^{t}dt'f_{l}(t')\cos(\Omega t')e^{i\omega_{\lambda} t'}P_{l}[\cos(-\Omega t' +\phi_{j}^{0})]  \nonumber\\
  g_{j,l}^{\lambda y} &\equiv& \sin(\Omega t)\int_{0}^{t}dt'f_{l}(t')\cos(\Omega t')e^{i\omega_{\lambda} t'}P_{l}[\cos(-\Omega t' +\phi_{j}^{0})] \nonumber \\
  h_{j,l}^{\lambda x} &\equiv& \cos(\Omega t)\int_{0}^{t}dt'f_{l}(t')\sin(\Omega t')e^{i\omega_{\lambda} t'}P_{l}[\cos(-\Omega t' +\phi_{j}^{0})] \nonumber \\
  h_{j,l}^{\lambda y} &\equiv&  \sin(\Omega t)\int_{0}^{t}dt'f_{l}(t')\sin(\Omega t')e^{i\Omega_{\lambda} t'}P_{l}[\cos(-\Omega t' +\phi_{j}^{0})]. \nonumber \\
   &&
\end{eqnarray}
It is possible that results drawn from these general formulas could be examined in benchmarking experiments such as the measurement of the spin-echo response or an average spin-spin interaction, when the laser beams have an effective wavevector parallel to the crystal plane.

Another interesting limit is the occasion in which the simulation time
$T$ is much shorter than the inverse of the rotational frequency
$\Omega T \ll 1$. In this limit,
the optical dipole interaction is given by
\begin{equation}
\label{eq:odp}
{\hat H}_{OD} = -\sum_{j=1}^{N}F_{O}\delta {\hat R}_{j}^{x}\sin(\delta k_{x}R_{j}^{0}\cos{\phi_{j}^{0}}-\mu t)\sigma_{j}^{Z},
\end{equation}
in which the relation $x_{j}^{0}(t)\approx R_{j}^{0}\cos{\phi_{j}^{0}}$ is used.
 By direct evaluation of the effective spin Hamiltonian in Eq.~(\ref{eq:spin_hamiltonian}),
 the effective Ising Hamiltonian ${\hat H}_{spin}$ can be cast into the form
 \begin{eqnarray}
   {\hat H}_{spin} &=& \hbar F_{O}^{2}\sum_{j j'\lambda \nu\nu'}{\rm Im}\Big[\alpha_{\lambda}^{\nu *}\alpha_{\lambda}^{\nu'}f_{j}^{\lambda}(t)\sin{(\phi_{j'}^{0}-\mu t)} e^{-i\omega_{\lambda} t}\Big]\nonumber\\
&\times&b_{j}^{x\nu}b_{j'}^{x\nu'}\sigma_{j}^{Z}\sigma_{j'}^{Z} \nonumber \\
    \ f_{j}^{\lambda}(t) &\equiv& \int_{0}^{t}dt'e^{i\omega_{\lambda}t'}\sin(\phi_{j}^{0}-\mu t').
 \end{eqnarray}
Since the contribution from fast oscillatory frequency components in $\hat{H}_{spin}$ is expected to be small because the contribution to the quantum phases are averaged out, we disregard the high frequency modes $\pm 2\mu$ under the rotating wave approximation. We arrive at the main conclusion that the effective spin simulator in the slow limit ($\Omega T \ll 1$) can be described by the static Ising model with the interaction $\hat{H}_{spin}=\sum_{jj'}J_{jj'}^{0}\sigma_{j}^{Z}\sigma_{j'}^{Z}$
in which the static Ising coupling is given by the expression
\begin{eqnarray}
  J_{jj'}^{0} &=& \sum_{\lambda \nu \nu'}\frac{\hbar F_{O}^{2}}{2(\mu^2-\omega_{\lambda}^2)}
\Big[\omega_{\lambda}\cos{\phi_{jj'}}{\rm Re}\{\alpha_{\lambda}^{*\nu}\alpha_{\lambda}^{\nu'}\} \nonumber \\
   &-& \mu\sin{\phi_{jj'}} {\rm Im} \{\alpha_{\lambda}^{*\nu}\alpha_{\lambda}^{\nu'}\}\Big]
   b_{j}^{x\nu}b_{j'}^{x\nu'},
\end{eqnarray}
and the angle $\phi_{jj'}=\phi_{j}^{0}-\phi_{j'}^{0}$ is determined by the ion equilibrium configuration. Because the function $\sin{\theta_{jj'}}$ is
odd for the permutation $j\leftrightarrow j'$ and the expression $\sum_{\nu\nu'}b_{j}^{x\nu}b_{j'}^{x\nu'}$ after the permutation $\nu\leftrightarrow\nu'$ is unchanged, only the first term in
$J_{jj'}^{0}$ contributes to the effective Ising Hamiltonian ${\hat H}_{spin}$ after the summation over the indices $j$ and $j'$. We therefore arrive at the main conclusion for the planar mode spin Hamiltonian $H_{spin}=\sum_{jj'}J_{jj'}^{0}\sigma_{j}^{Z}\sigma_{j'}^{Z}$ at slow rotation $wT\ll1$ where the spin-spin interaction is described by
\begin{equation}
J_{jj'}^{0} = \sum_{\lambda \nu \nu'}\frac{\hbar \omega_{\lambda} F_{O}^{2} {\rm Re}\{\alpha_{\lambda}^{*\nu}\alpha_{\lambda}^{\nu'}\} }{2(\mu^2-\omega_{\lambda}^2)}
   b_{j}^{x\nu}b_{j'}^{x\nu'}\cos{\phi_{jj'}}.
\label{eq:PlanarJij}
\end{equation}
The effects
of the distribution of the ion equilibrium positions from the interaction
in Eq.~(\ref{eq:odp}) can be interpreted as a varying phase fluctuation due to the external symmetry breaking from the laser beams. In our later numerical discussion, we show two planar mode branches. One branch has much lower energy than the other branch. The quantum simulation with phonons in the lower branch is particularly relevant for experimental realization when the  laser is closely detuned from that branch. In this case, the larger spin-spin exchange interaction (which is inversely proportional to ${\mu^2-\omega_{\lambda}^2} \approx 2\omega_{\lambda}\times(\mu-\omega_{\lambda})$ where $\mu \approx \omega_{\lambda}$ guarantees that the simulation can be realized at a much shorter time scale before the decoherence from the environment takes place.

\section{Numerical results}
We have discussed above the theoretical formulation for the description of phonon-mediated quantum spin dynamics.
Due to the inhomogeneity and finiteness of the experimental systems, numerical studies
are necessary to understand the behavior and facilitate a detailed understanding of current and future cold-ion experiments in Penning traps.

\subsection{Typical system parameters}
We consider  ${}^{9}\textrm{Be}^{+}$ ions trapped in a static magnetic
field with a rotating wall potential $V_{W} r^{2} \cos2(\theta +\Omega t)$ ($\Omega > 0$). The axial trapping frequency due to the end-cap potentials $eV_{0}=\frac{1}{2}m\omega_{z}^2$ is fixed at the value $\omega_{z}=2\pi \times 795$ kHz (in units of rad/s) as typically applied in experiments. We measure all angular frequencies in units of $\omega_z$. For example, we define the frequency scale
$\omega_W$ associated with the rotating wall potential $V_{W}$
by the relation $\omega_W = \sqrt{2e|V_{W}|/m}$.
We choose cases with very weak, weak, and strong rotating wall
potentials given by the corresponding values $\omega_W= 0.01\omega_{z}, 0.04 \omega_{z}$, and $0.07  \omega_z$.
The cyclotron frequency $\omega_{c}$ associated with the magnetic field is $\omega_{c}=eB_{z}/m$. With the magnetic field $B_{z}=4.5$ Tesla, the value of the cyclotron frequency is given by $\omega_{c}=9.645 \omega_z$.
The Beryllium ion has an atomic mass $m = 9.012182$u,
where u is the atomic mass unit, and has a positive unit charge
$e= 1.60217646\e{-19}$ Coulomb.

\begin{figure}[htbp!]
  \centering
    \includegraphics[width=0.45\textwidth]{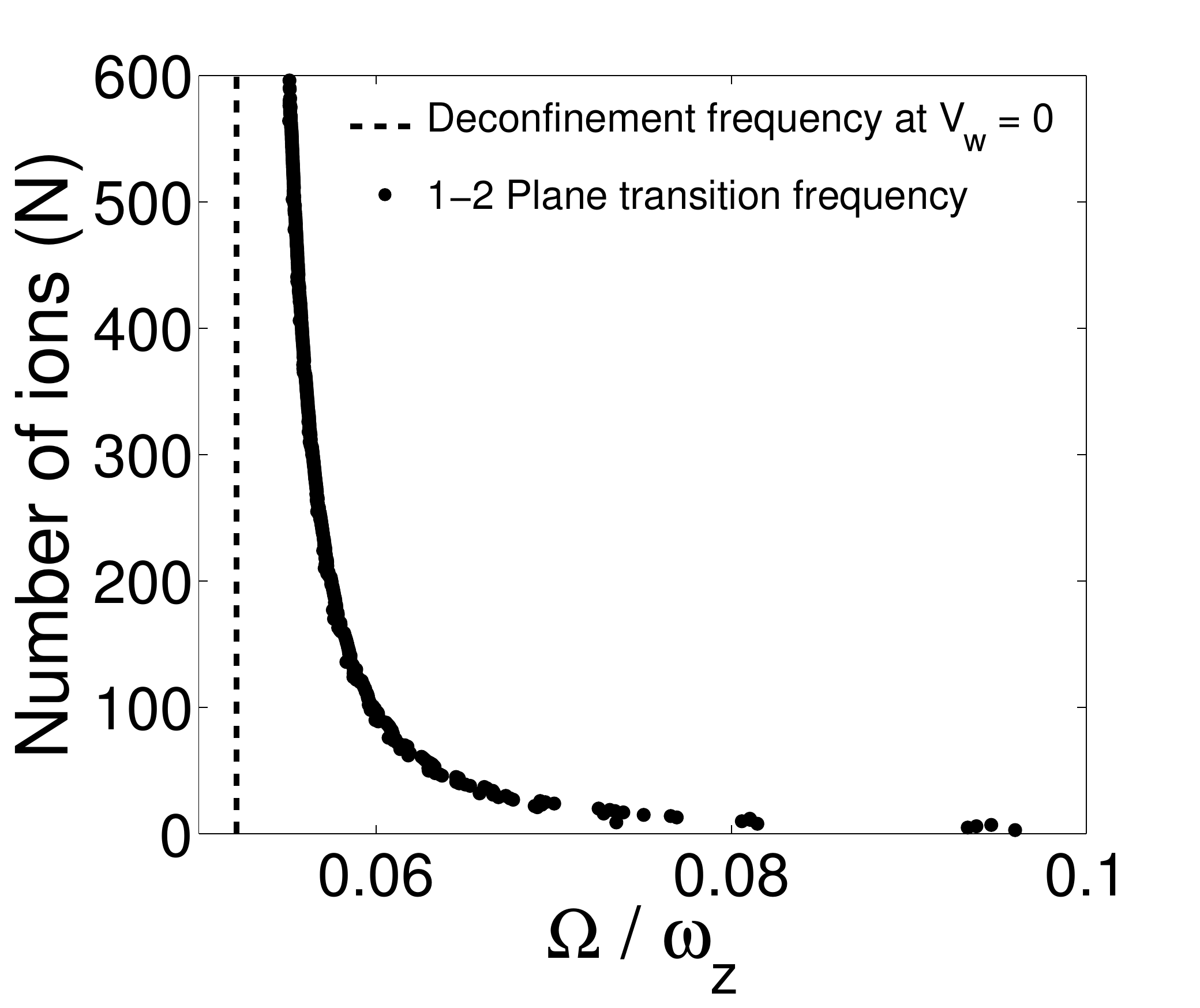}
\caption{One-to-two plane transition frequency for different numbers of ions $N$. The vertical dashed line shows the deconfinement frequency for $V_{W} = 0$.
For larger values of $V_{W}$, the vertical line moves to the right, reducing the frequency range for the rotating wall, where the single plane is stable, while the one-to-two plane transition line hardly changes.}
\label{PT12}
\end{figure}

Based on our theoretical discussion above, the system is energetically stable when
the rotational frequency $\Omega$ due to the rotating wall potential lies
between the deconfinement transition frequency $\omega_{dc}$ and the one-to-two
plane transition frequency $\omega_{12}$.
The deconfinement frequency $\omega_{dc}$ is determined by the criterion
$\beta_{3}=0$. This deconfinement criterion can be recast into the following form:
\begin{equation}
\omega_{dc}= \frac{\omega_c}{2} - \sqrt{\frac{\omega_c^2}{4} - \frac{\omega_z^2}{2} -\omega_W^2}.
\end{equation}
The one-to-two plane transition frequency $\omega_{12}$ needs to be reliably
determined numerically by the instability of the axial phonon frequencies.
We can map out the one-to-two plane instability by calculating when the axial phonon frequencies first become imaginary,
as shown in Fig.~\ref{PT12} for $V_W=0$.  The one-to-two plane transition frequency depends only weakly on the wall potential via the change of the equilibrium coordinates due to the presence of the wall.

In our numerical discussion, we choose three rotation frequencies
$\Omega=\omega_{12}-\Delta$,\ $\Omega=0.5(\omega_{dc}^{0}+\omega_{12})$,\
$\Omega=\omega_{dc}+\Delta$[$\omega_{dc}$ depends on  $V_{W}$] with the frequency offset $\Delta=2\pi \times 200$ Hz where the frequency $\omega_{dc}^{0}$ is the deconfinement frequency at zero rotating wall potential ($V_{W}=\omega_W=0$).
Equivalently, instead of the rotational frequency, we report the effective trapping frequency in the crystal plane
$\omega_{eff} = \sqrt{\omega_c \Omega - \Omega^2 - \frac{eV_0}{m}}$
where the relation $eV_{0}=0.5m\omega_{z}^{2}$ is used.
Thus, we can report the corresponding scaled effective frequency $\omega_{h} = 0.21\omega_z$ for the high rotational frequency case $\omega_{12}-\Delta$,
the corresponding mean effective trapping frequency $\omega_{m} = 0.16 {\omega_z}$ for the mean rotational frequency case $0.5(\omega_{dc}^{0}+\omega_{12})$,
and the low effective trapping frequencies for the low frequency case $\Omega=\omega_{dc}+\Delta$ with very weak, weak and strong rotating wall potentials as
$\omega^{VW}_{l} = 0.05\omega_{z}$, $\omega^W_{l} = 0.06 {\omega_z}$,
and $\omega^S_{l} = 0.09 \omega_z$, respectively. We have three cases for the low frequency case because the deconfinement frequency depends strongly on the wall potential near the low frequency, so it must change with the wall potential; for the other two cases the dependence is so small that we ignore it.

\subsection{Equilibrium configurations}

To seek for the equilibrium configuration of crystallized ions in the trap, we generate a seed lattice of finite size, which takes the form
of a triangular lattice near the center and has the boundary as a closed hexagonal shell, with six facets.  We look for a stable equilibrium configuration which
minimizes the Hamiltonian in the rotating frame, assuming none of the ions are moving. We find this procedure always guarantees that we find a local minimum that looks like a triangular lattice in the center, just as is found in experiment but has an elliptical boundary which is shaped by the equipotential of the effective trapping potential (this approach appears to be similar to that in Ref.~\onlinecite{Chinese}, but we use a different potential by including the rotating wall, while it is different from the approach used in Ref.~\onlinecite{Jake}). The minimization procedure employs both a locally calculated gradient and a locally calculated Hessian matrix, similar to the method described in Ref.~\onlinecite{peeters}. We used the MatLab optimization toolbox and our own code to calculate these minimal configurations, and results between the two codes agreed very well.

\begin{figure}[htbp!]
  \centering
    \includegraphics[scale=0.45]{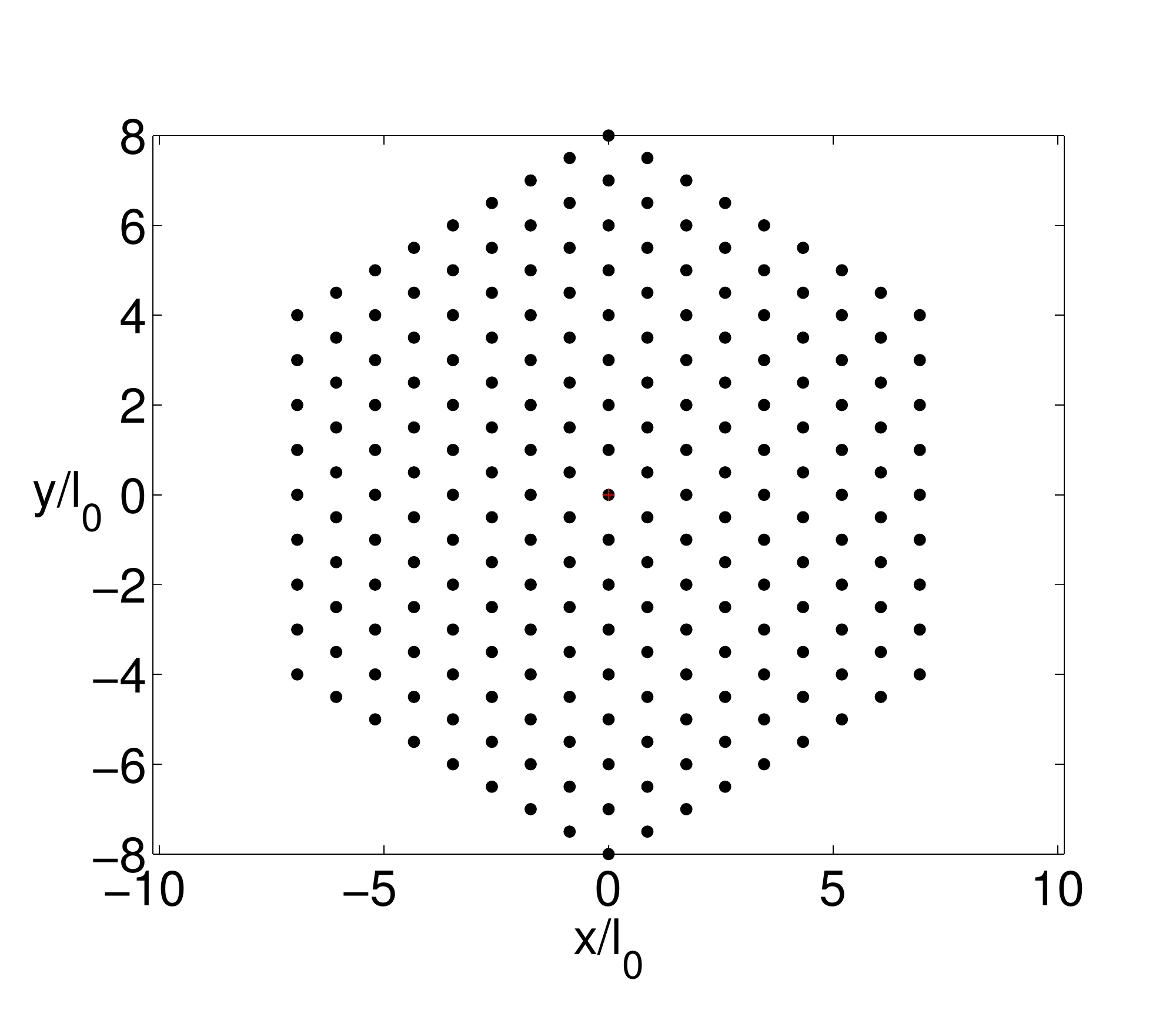}
\caption{Spatial configuration for a seed lattice with $N = 217$ ions. The spatial dimension is scaled by the characteristic length in the axial direction ${\hat z}$ as $l_0 = \left(\frac{k_{e}e}{V_0}\right)^{\frac{1}{3}}=\left(\frac{2k_{e}e^2}{m\omega_{z}^2}\right)^{\frac{1}{3}}$ The red cross denotes the point where the external potential due to the electrodes vanishes. The minimal ion configuration often, but not always, has an ion located at this energy minimum. }
\label{seed}
\end{figure}

The construction of the seed lattice is based on a closed shell construction
analogous in spirit to closed-shell electron configurations for atoms.
More specifically,
for any particle number such as $N=217$, we first determine how many
closed hexagonal shells can be created with that number of ions.
The lattice is a closed hexagonal lattice if the
boundary of the initial seed lattice forms a perfect hexagonal shape.
The number of the closed hexagonal shells $S$ is related to the number
 of ions $N$ for the seed lattice by
\begin{equation}
S = \left\lfloor\frac{\sqrt{9+12(N-1)}-3}{6}\right\rfloor.
\end{equation}
 If the number of ions $N$ cannot fill an integer number of hexagonal shells, the outermost ions are placed on an incomplete outer hexagonal ring according to the minimal potential energy at each of the outer ring sites (the energies are different because the outer ring is not elliptical in shape, especially when $V_W$ is nonzero). For the case with $N = 217$,
an integer number of closed hexagonal shells are generated in the seed lattice ($S=8$) which is shown in Fig.~\ref{seed}.

\begin{figure}[htbp!]
\centering
\hspace*{-0.4in}
\includegraphics[scale=0.3]{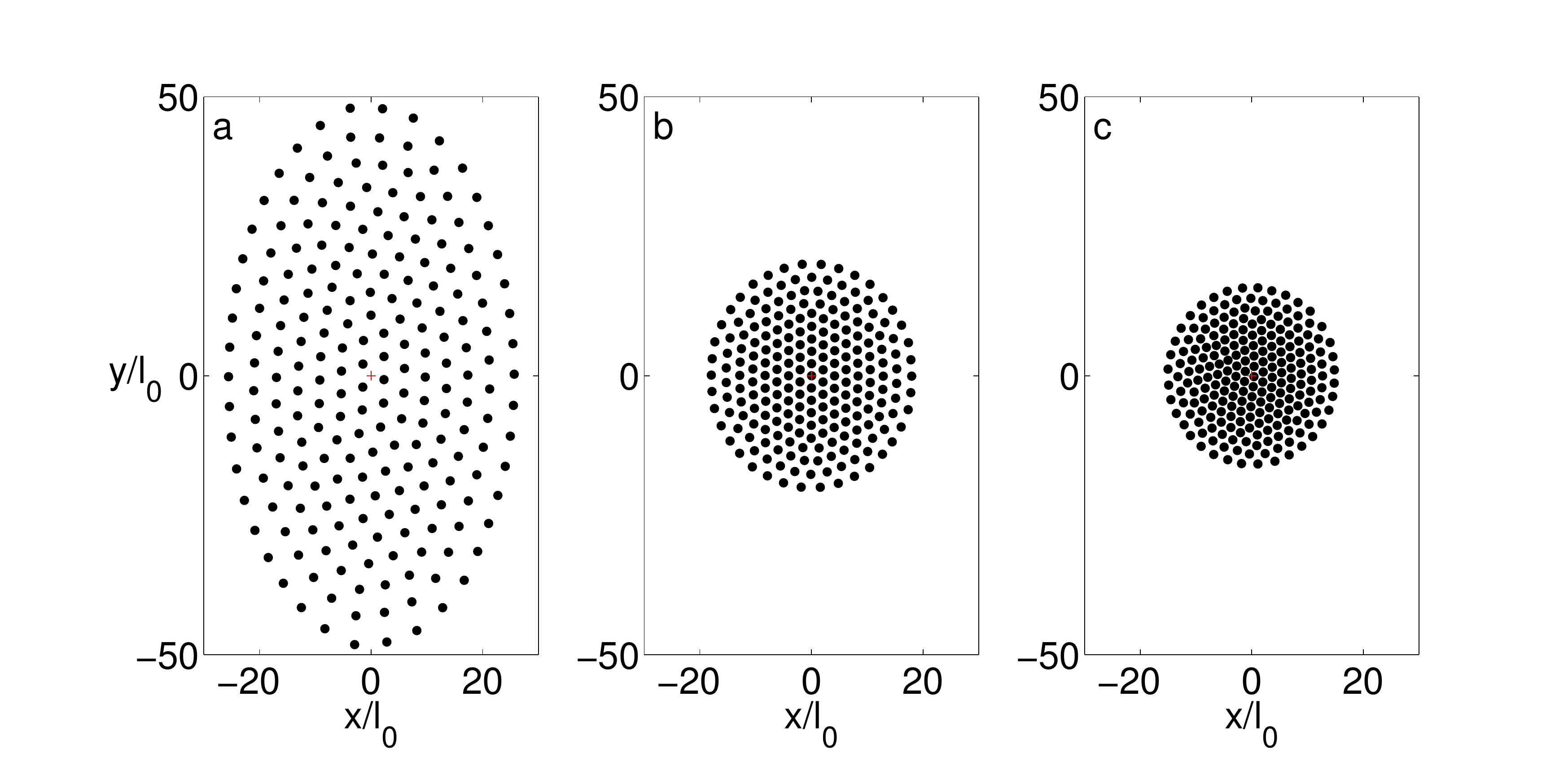}
\caption{Equilibrium structures for different rotating frequencies: (a) Low effective trapping frequency $\omega^W_{l}=0.06 \omega_{z}$.
(b) Medium effective trapping frequency $\omega_{m}=0.16 \omega_{z} $. (c) High effective trapping frequency $\omega_{h}=0.21 \omega_{z}$. The wall potential is $\omega_W=0.04\omega_z$ for all cases. }
\label{EqPos}
\end{figure}

In Fig.~\ref{EqPos}, we show spatial equilibrium configurations at different rotational frequencies with the same fixed weak rotating potential $\omega_{W}=0.04\omega_z$.  We expect that the anisotropy introduced by the rotating wall
potential $\omega_{W}$ is minimal when the effective trapping potential dominates
over the  rotating wall potential $\omega_{W}$.
As shown in the Fig.~\ref{EqPos}(c), we clearly observe that the ions are arranged much more
isotropically with a much higher density than Fig.~\ref{EqPos}(a) and Fig.~\ref{EqPos}(b) due to the large effective trapping in the plane. Note how the seed lattice smoothly evolves from a nice triangular lattice in the center to a boundary that takes the minimal, elliptical shape, of the potential at the edges, in the stable equilibrium positions.

\begin{figure}[htbp!]
  \centering
    \includegraphics[width=.5\textwidth]{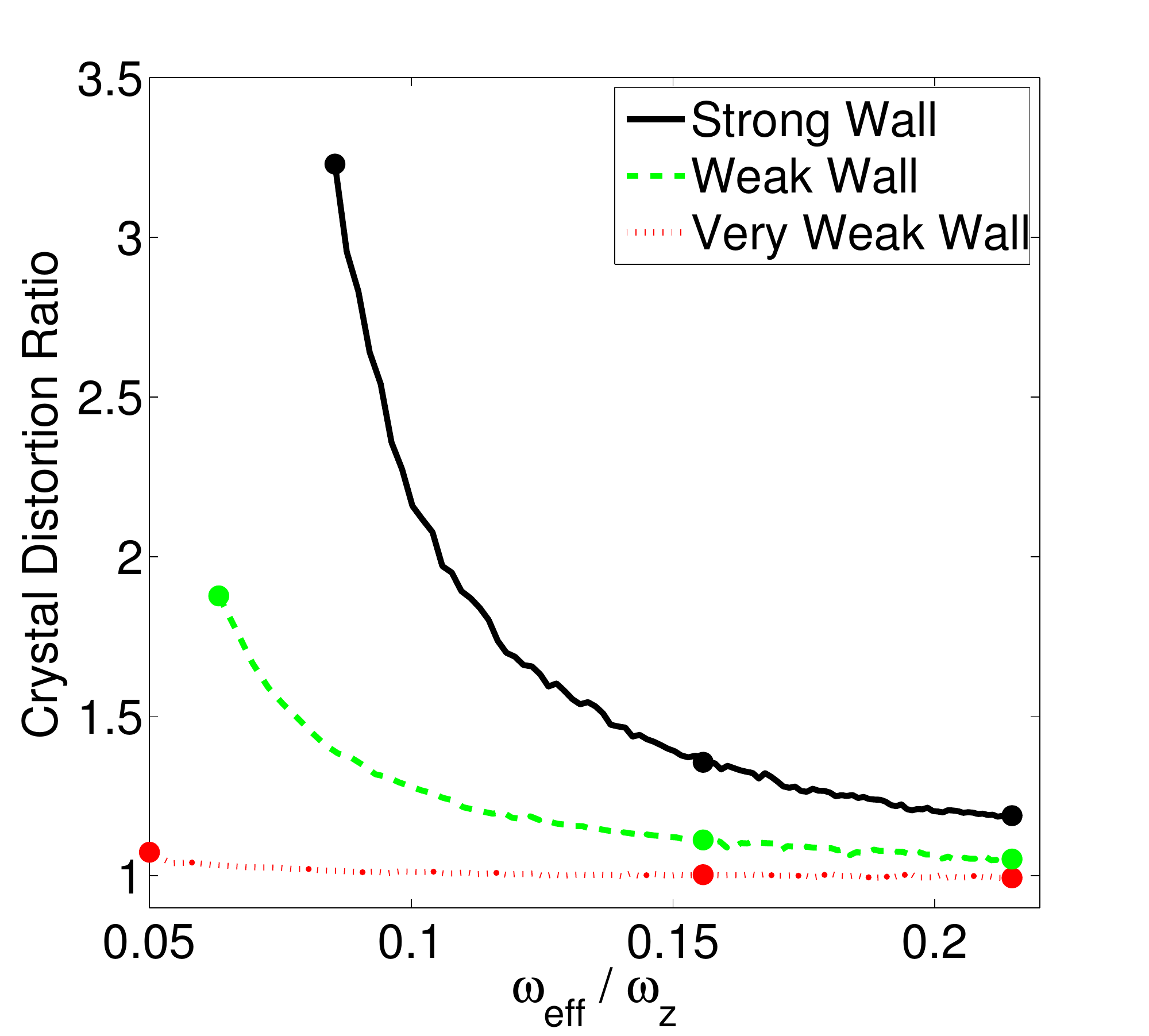}
\caption{(Color online.) Dependence of the crystal distortion ratio (or aspect ratio) on the rotational frequency $\omega_{eff}$ and the rotating wall potential $\omega_{W}$.
The black-solid curve represents a strong rotating wall with $\omega_{W}=0.07\omega_{z}$. The green-dashed curve
is a weak rotating wall potential with $\omega_{W}=0.04\omega_{z}$. The red-dotted curve is a very weak
rotating wall potential with $\omega_{W}=0.01\omega_{z}$.
The colored dots are the principle cases we examine with more detailed calculations for much of the remaining numerical work.
}
\label{distort}
\end{figure}

To further quantify the change in isotropy, we define the distortion of the crystal as the aspect ratio of the major axis to the
minor axis of the elliptical edge. Fig.~\ref{distort} shows the dependence of this distortion on the rotational frequencies
$\omega_{eff}$ and rotating wall potential $\omega_{W}$.
With a fixed rotating wall potential, we observe the lattice is more distorted with a weaker effective frequency $\omega_{eff}$. The effect is much larger for the strong rotating wall potential $\omega_{W}$.
This shows that the role of the rotating wall potential  is not always negligible. Its effect can only be neglected when the ratio between the potential and effective trapping frequency $\omega_{W}/\omega_{eff}$ is
smaller than one.

\begin{figure}[htbp!]
  \centering
    \includegraphics[width=.5\textwidth]{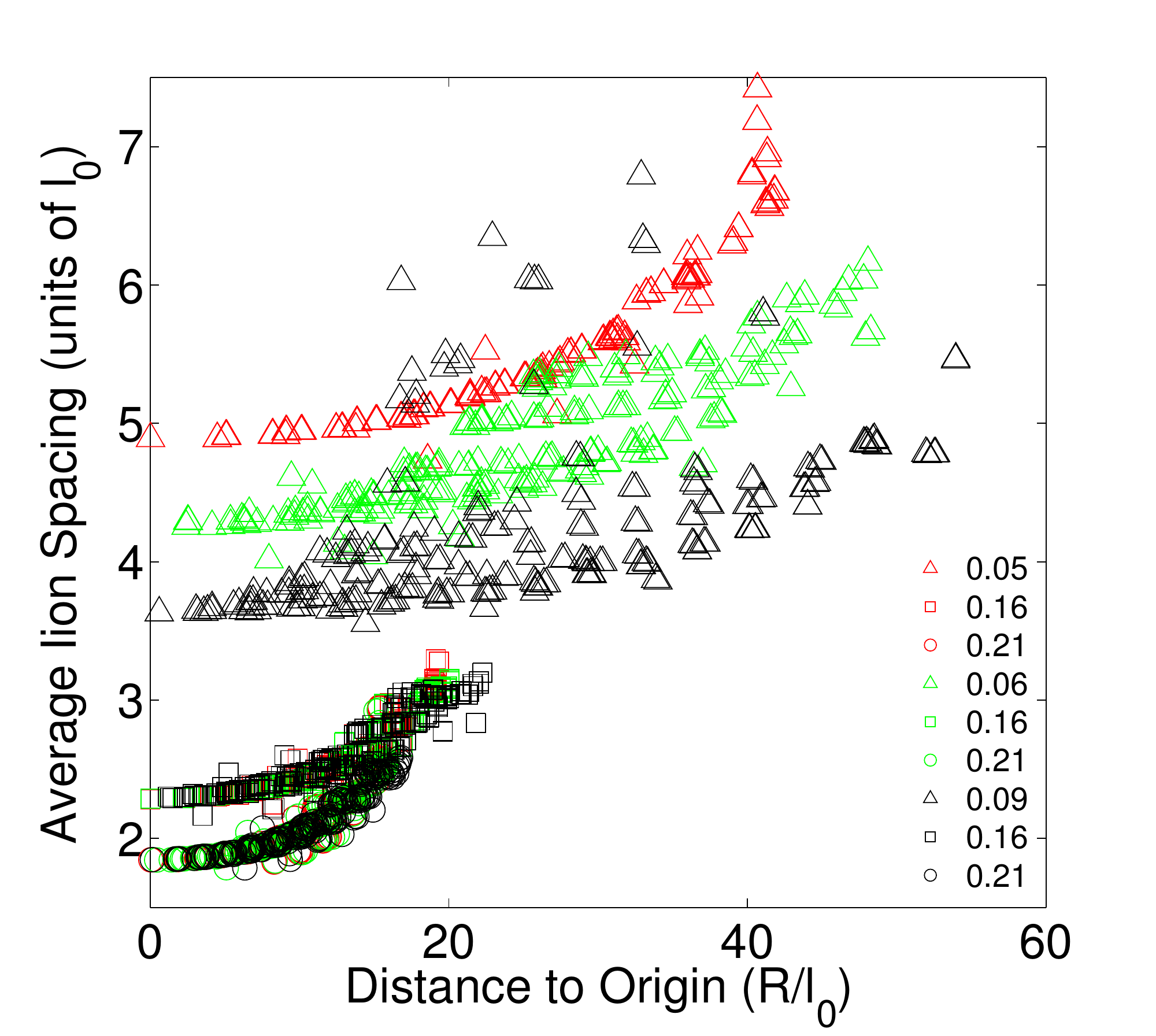}
\caption{Ion spacing dependence as a function of the distance $R$ from the origin. The black, green, and the red symbols denote respectively the data with a  strong rotating wall potential $\omega_{W}=0.07\omega_{z}$, a weak rotating wall potential $\omega_{W}=0.04\omega_{z}$, and a very weak rotating wall potential $\omega_{W}=0.01\omega_{z}$. The values of the corresponding  effective trapping frequencies $\omega_{eff}$ are shown in the legend. The same parameter sets as in Fig.~\ref{distort} are used.
}
\label{avgdist}
\end{figure}

Fig.~\ref{avgdist} shows how the ion spacing of the equilibrium
structure changes as a function of the distance from the origin. To compute the lattice spacing for a given ion, we find its nearest neighbors using a Delaunay triangulation algorithm and then calculate the mean distance from that ion to those nearest neighbors. Each data point represents the mean nearest-neighbor ion spacing for each ion at a different radius $R$.
When we have a strong rotating wall and a weak effective potential, we expect the ions are more
widely spread out and are pushed more along the weakly confined direction. This can be seen in
the data sets with colored triangles, which have a larger mean nearest-neighbor distance and a larger spread about the average.
For the data sets with colored squares and circles (weaker rotating wall potential), we observe much smaller mean nearest-neighbor distances with much less spread.

\begin{figure}[h!]
  \centering
    \includegraphics[width=.5\textwidth]{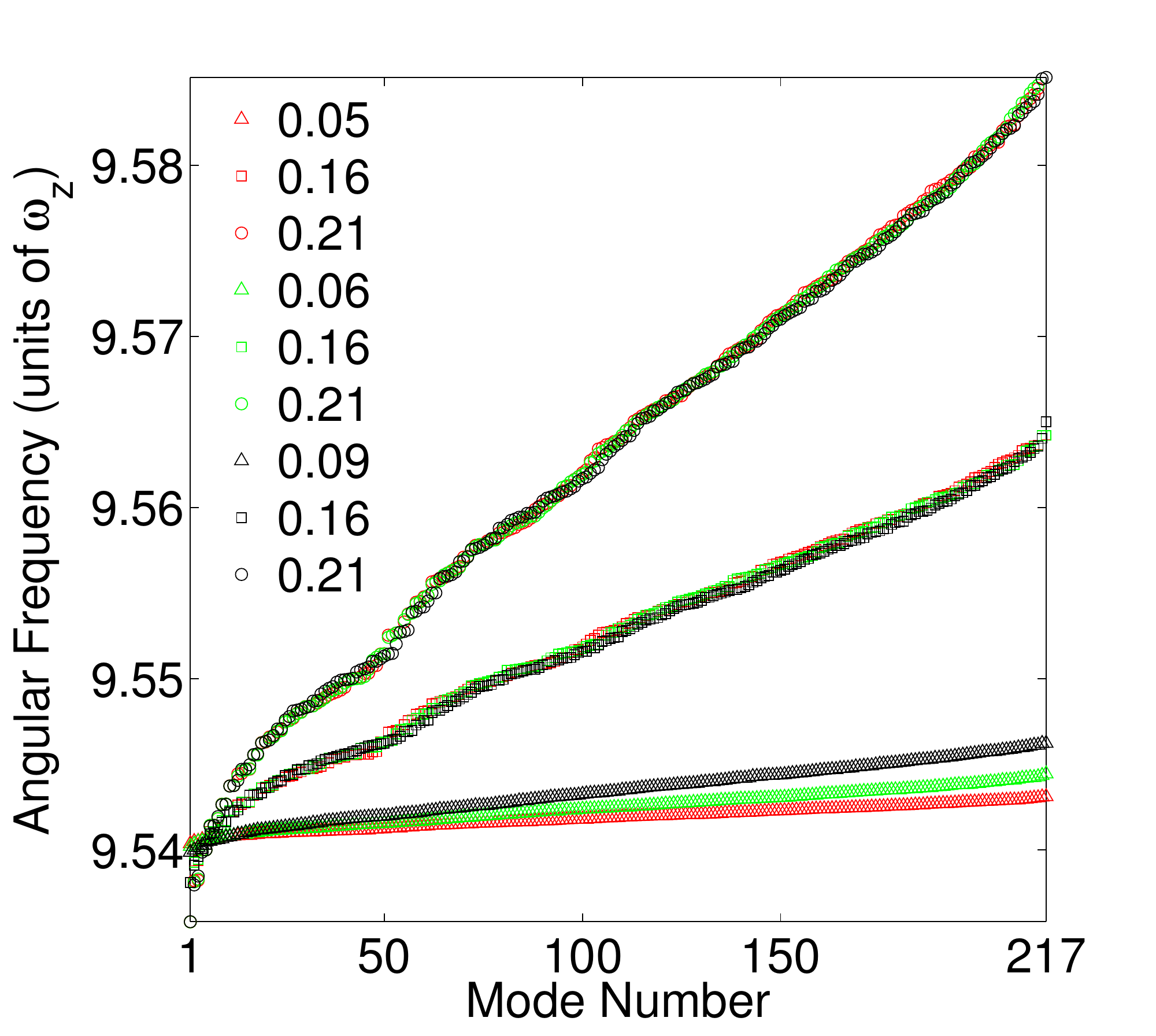}
\caption{(Color online.) Eigenfrequencies of the upper (cyclotron-like) branch of the planar modes.
The same parameter sets and notation are used as in Fig.~\ref{distort}.
}
\label{Upper}
\end{figure}

\subsection{Normal modes}
\begin{figure}[h!]
  \centering
    \includegraphics[width=.5\textwidth]{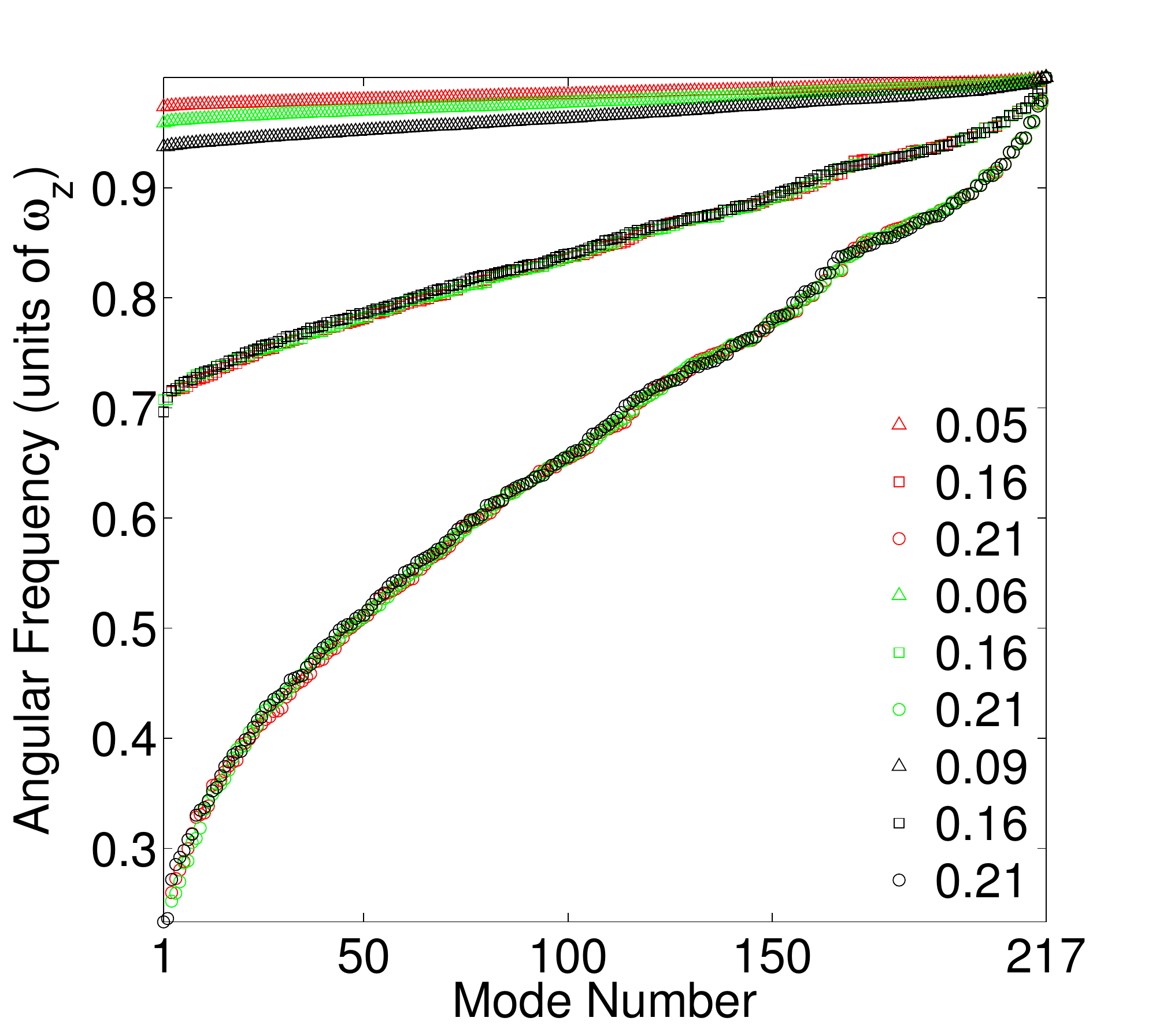}
\caption{(Color online.) Eigenfrequencies of the axial modes.
The same parameter sets and notation are used as in Fig.~\ref{distort}.}
\label{Axial}
\end{figure}

\begin{figure}[h!]
  \centering
    \includegraphics[width=.5\textwidth]{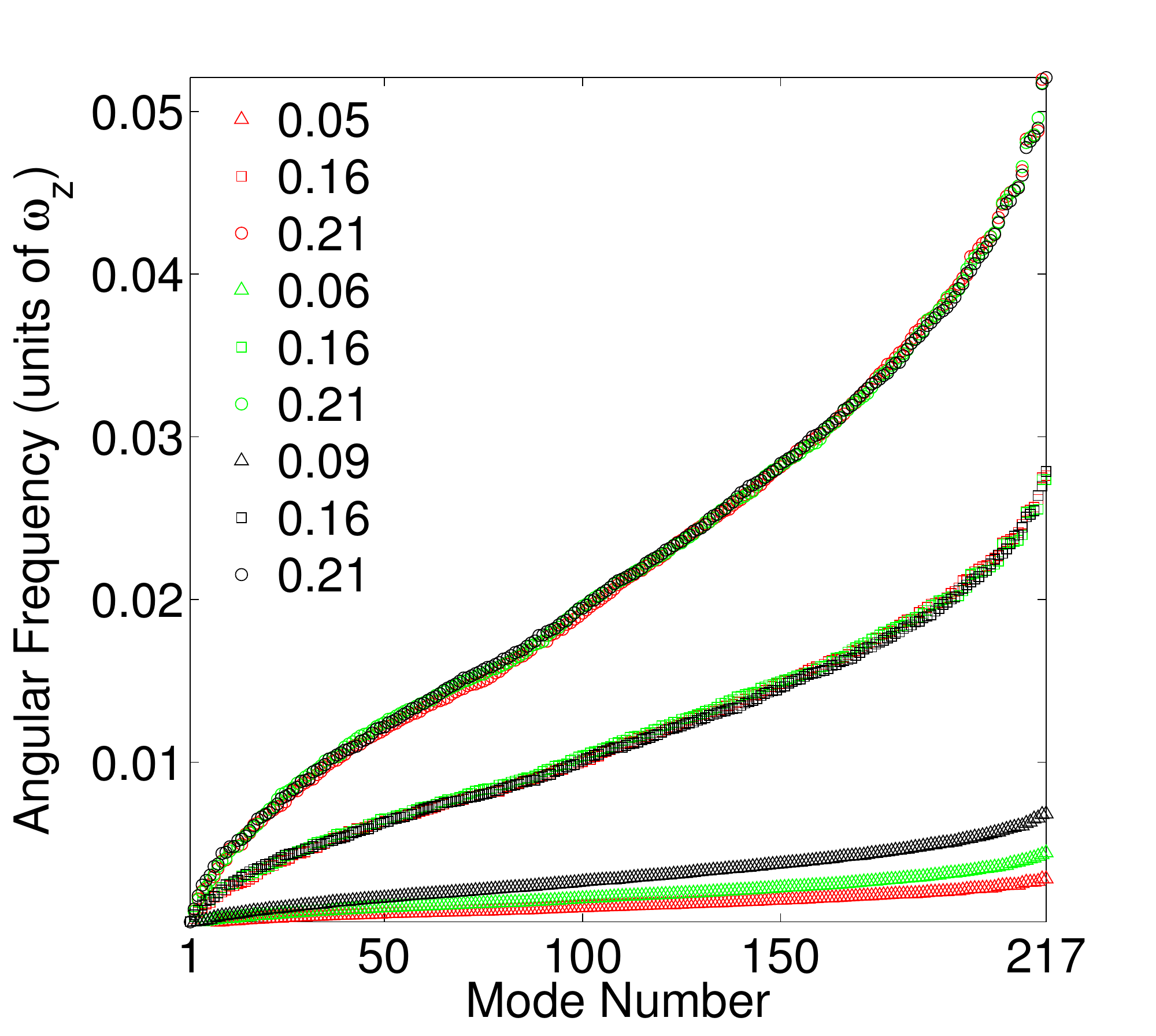}
\caption{(Color online.) Eigenfrequencies of the lower (magnetron-like) branch of the planar modes. The same parameter sets and notation are used as in Fig.~\ref{distort}.}
\label{Lower}
\end{figure}

In this subsection, we discuss the nature of the normal modes including the energy spectrum and properties of the eigenvectors.
Here we present the eigenfrequencies from each of the three phonon branches (one branch from the axial degrees of freedom and two branches from the planar degrees of freedom). We find in general  that the three branches are well separated in energies with the bandwidth strongly dependent on the equilibrium configuration of the ions.
The upper planar modes (shown in Fig.~\ref{Upper}) have the character of the high frequency cyclotron motion renormalized by the Coulomb interaction
between ions and the lower planar modes (shown in Fig.~\ref{Lower}) have the character of the low frequency magnetron motion. The eigenfrequencies of the axial mode branch shown in (Fig.~\ref{Axial}) lie in between the two planar mode branches. In each figure, the modes are sorted by ascending frequency. In general, eigenfrequencies depend mostly on the effective frequency $\omega_{eff}$ and are less sensitive to the small perturbation
due to the rotating wall potential $\omega_{W}$ even though the character
of the eigenvectors can be quite sensitive to the rotating wall potential.

In Fig.~\ref{Upper}, the dependence of the cyclotron-like eigenfrequencies on the rotating wall potential and rotational frequency is shown for the nine typical cases we will be examining in our numerical results.
Cases are represented by the same color for the same $\omega_{W}$ with the same symbol shape for the same $\omega_{eff}$. Note how it is the effective frequency that primarily determines the frequency spectrum, with a separation due to the rotating wall potential only becoming clearly visible when that potential is strong.
In addition, note the lowering of the lowest eigenfrequency with
large $\omega_{eff}$, which signals a collective feature due to Coulomb interactions (colored circles).

In Fig.~\ref{Axial}, we observe one axial mode (the highest one) which has a universal
eigenfrequency with $\omega=\omega_{z}$ independent of the parameter sets. This is due to the fact that the uniform center of mass motion along the axial direction does not cost any Coulomb energy and has the same trapping energy in the axial direction so it is independent
of the rotating wall potential applied in the crystal plane.
This is shown as the intersection of the highest axial modes for all the parameter sets at the highest axial frequency ($\omega=\omega_{z}$). The other phonon modes have lower eigenfrequencies than
the center of mass (COM) mode. This can be interpreted as the reduction of the Coulomb interaction due to the increase of the average ion spacing from the phonon displacement when the wavevector is nonzero (for the COM mode, the phonon does not change the relative displacement of the ions, so the Coulomb interaction is unchanged). This interpretation also agrees with the observation that the axial mode branches for large $\omega_{eff}$ lie lower in energy as most explicitly shown with the colored
circle data. The phonon frequencies near the COM mode have been verified against experimental measurements using a technique that can measure the effective temperature per phonon mode and the actual frequency of each phonon mode~\cite{Brian}.

In Fig.~\ref{Lower}, we plot the lowest (magnetron-like) planar modes for the same nine cases. When there is no rotating wall potential, the lowest-frequency mode is that of a rigid rotation, which costs no energy since the potential in the plane depends only on the radius in that case.
This zero energy mode is the Goldstone mode due to the spontaneous rotational symmetry breaking of the equilibrium configuration. As we break the rotational symmetry by turning on a wall potential, the Goldstone mode ``acquires mass" and becomes nonzero, but often remains a very low frequency mode with a near rigid rotation character.
The bandwidth of this branch of planar modes is quite sensitive to the effective trapping frequency $\omega_{eff}$ as can be shown by examining the the colored circle and colored square cases. The bandwidth is only slightly
dependent on the rotating wall potential as can be seen with
the colored triangular cases.

\begin{figure}[h!]
  \centering
    \includegraphics[width=.5\textwidth]{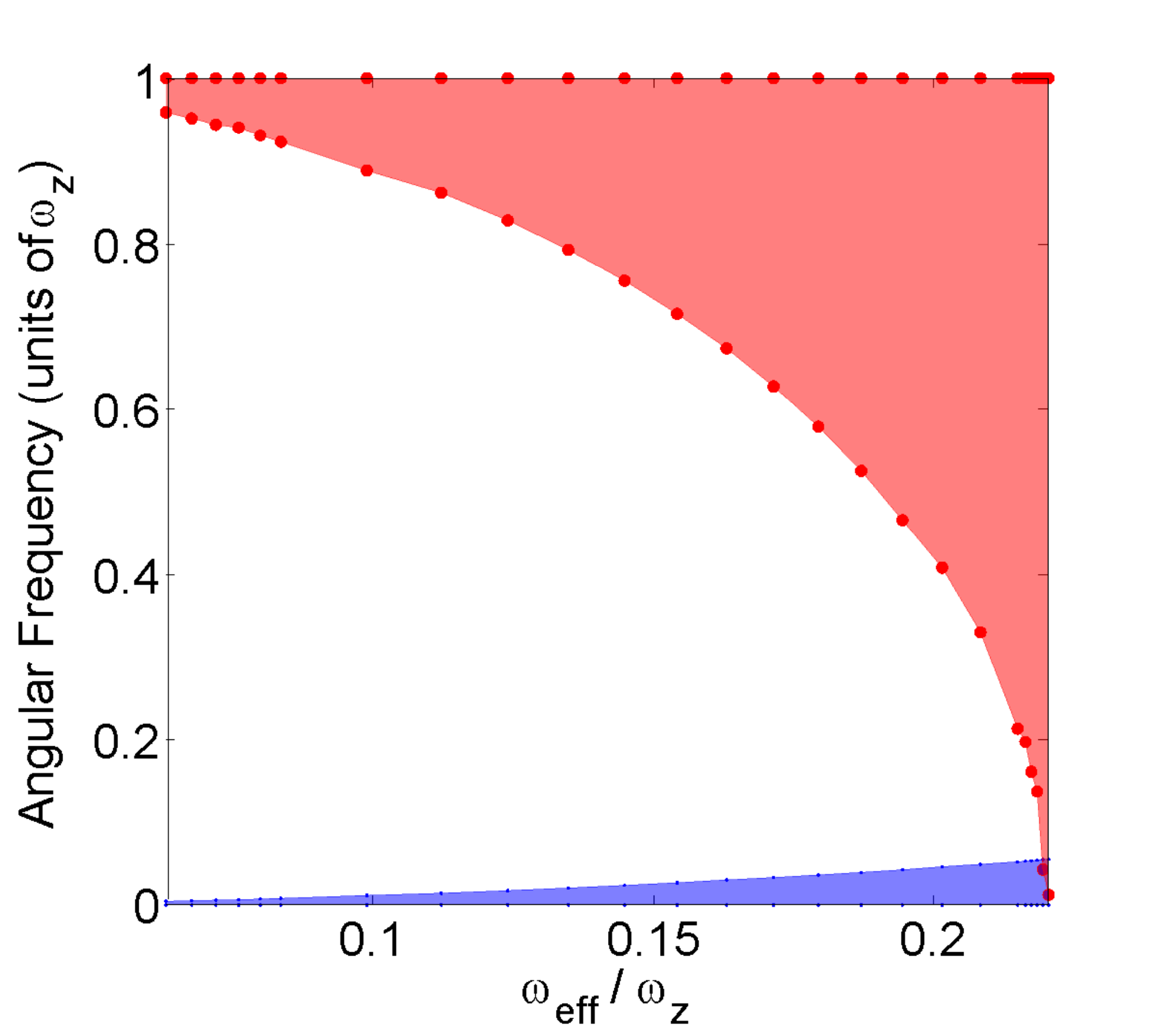}
\caption{(Color online.) Comparison of the phonon bandwidths for the axial (red) and lower (magnetron-like) planar (blue) branches for a fixed weak rotating wall potential.
As the effective trap frequency increases, the lowest frequency of the axial mode decreases
as the largest (magnetron-like) lower branch planar frequency increases until they overlap
very near the one-to-two plane transition. The uppermost axial mode is fixed, while the lowest planar mode does not change significantly with $\omega_{eff}$.}
\label{Overlap}
\end{figure}

As the rotational frequency is increased relative to the axial frequency, the confinement in the plane becomes tighter and tighter until, suddenly, the system becomes stabilized in a two-plane configuration. As we approach this critical frequency, the axial mode phonon bandwidth grows tremendously, as the lowest axial frequency marches down and eventually becomes lower than the maximal magnetron-like planar phonon frequency (which increases with the rotational frequency but not as rapidly), as shown in Fig.~\ref{Overlap}. The overlap occurs (at $\omega_{eff}=0.2189\omega_z$) quite close to the critical one-to-two-plane transition ($\omega_{eff}=0.2202\omega_z$) for a rotating wall potential corresponding to $\omega_W=0.04\omega_z$.

\begin{figure}[htbp!]
  \centering
    \includegraphics[width=.48\textwidth]{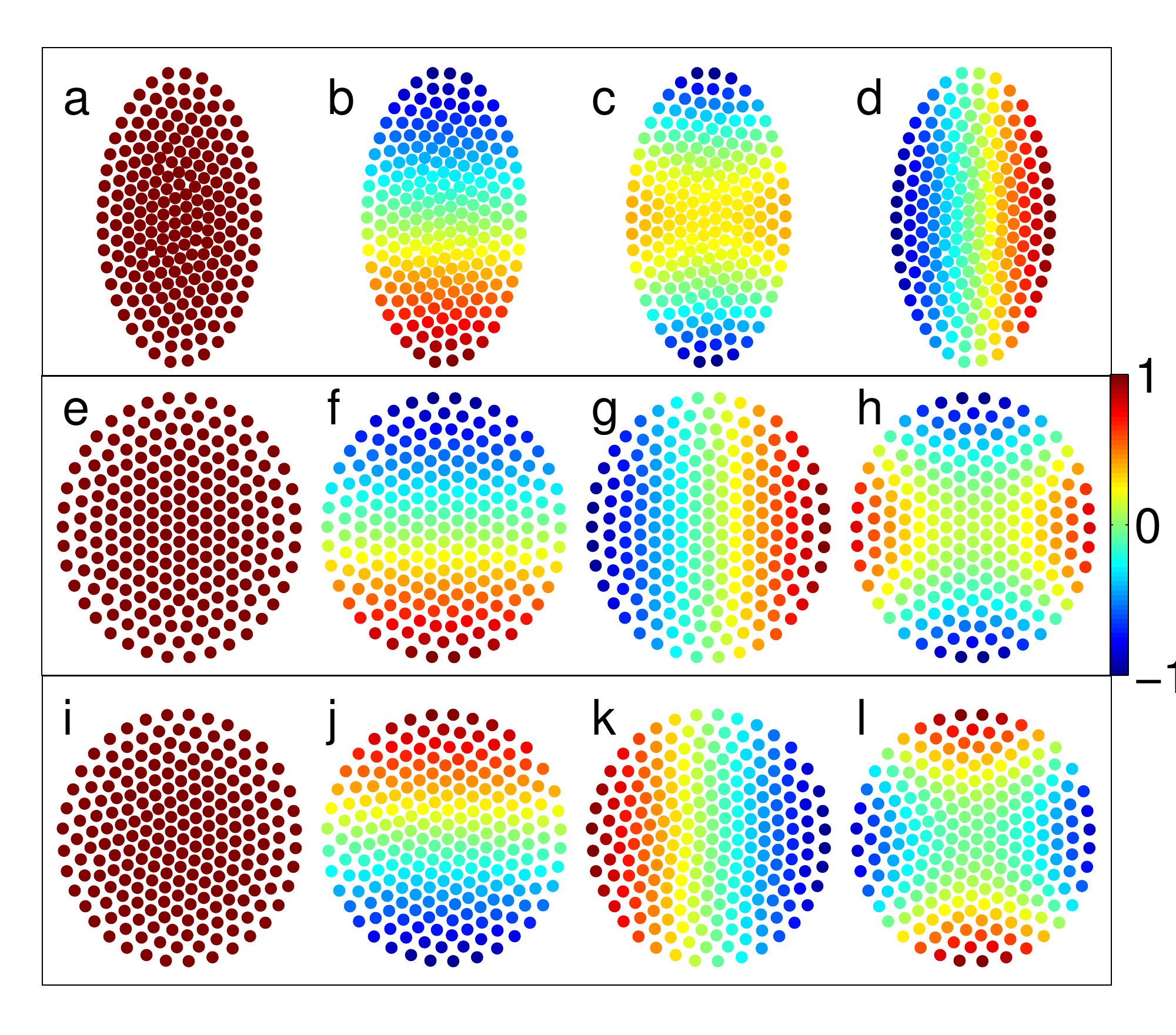}
\caption{(Color online.) Four highest frequency axial eigenvectors for three different effective trapping frequencies with a weak rotating wall potential $\omega_{W}=0.04\omega_{z}$.
Subplots (a)-(d) present the case with a low effective trapping frequency $\omega^W_{l}=0.06\omega_{z}$, (e)-(h) present the case with a mean effective trapping frequency $\omega_{m}=0.16\omega_{z}$, and (i)-(l) represent the case with a high effective trapping frequency $\omega_{h}=0.21\omega_{z}$. The color in the colorbar
represents the scaled value of the normalized eigenvector $b_{j}^{\nu z}$
with respect to its maximal positive component for the mode $\nu$.}
\label{CompareAxial}
\end{figure}

\begin{figure}[htbp!]
  \centering
    \includegraphics[width=.5\textwidth]{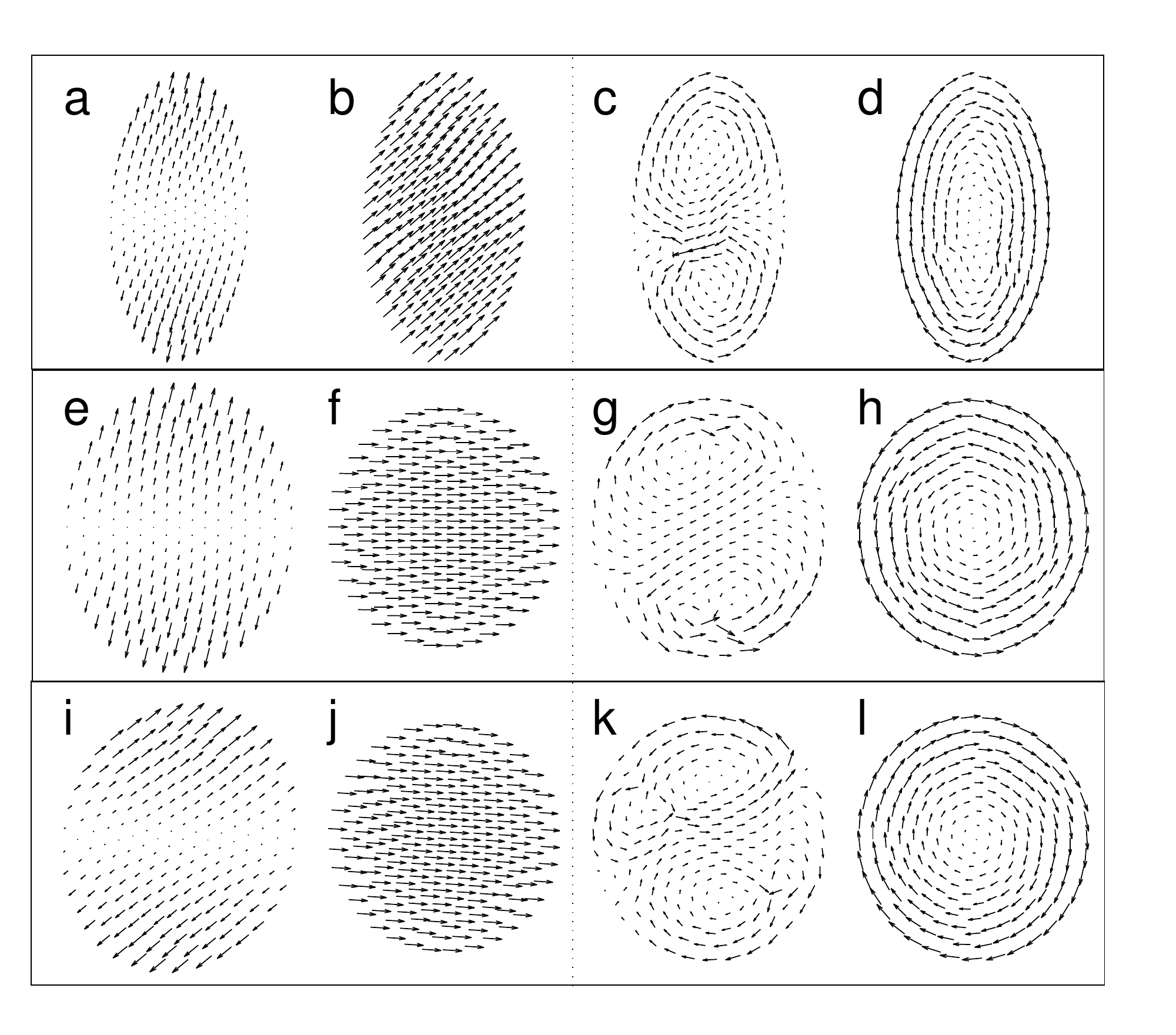}
\caption{A snapshot of the spatial eigenvectors for the lowest two modes in each planar branch at the fixed rotating wall potential $\omega_{W}=0.04\omega_{z}$. Subplots (a)-(b), (e)-(f), and
(i)-(j) show  the lowest two planar modes in the upper branch with the
effective trapping frequencies $\omega_{l}^{W}=0.05\omega_{z}$, $\omega_{m}=0.06 \omega_{z}$, and $\omega_{h}=0.09\omega_{z}$ respectively. Subplots (c)-(d), (g)-(h), and (k)-(l) on the right panel are the corresponding data for the lower branch of the planar phonon modes.}
\label{ComparePlanar}
\end{figure}

We now discuss the nature of the eigenvectors for the different  phonon modes. In general, we find phonon modes to either be collective, where all of the ions move with a long-wavelength, similar to the drumhead modes of the continuum, or to be localized, where the motion is confined primarily to a small fraction of the ions.  It is the long-wavelength modes that are most important for quantum simulation, so we focus on them.  For the axial branch, these modes lie close to the COM mode (at high frequency), while for the magnetron-like branch, they are close to the rigid rotation mode, at low frequency.
In Fig.~\ref{CompareAxial}, we show the four
highest frequency modes of the axial branch and in Fig.~\ref{ComparePlanar}, we show the two lowest
frequency modes of the upper (cyclotron-like) branch and lower (magnetron-like) branch planar modes, all for the same parameters as in
Fig.~\ref{EqPos}. Unlike Fig.~\ref{EqPos}, our focus is on the character of the eigenvectors instead of the absolute size of the equilibrium configurations, so we have rescaled the positions to make the ion positions have similar widths in all panels. Similarly, the color scales used for the axial modes and the arrow lengths used for the planar modes are both adjusted for each panel to highlight the details of the motion and should not be used to compare relative ion displacements from one panel to another.

We start with the axial modes in Fig.~\ref{CompareAxial}. The leftmost mode is the COM mode which involves uniform motion of all of the ions for all cases.  If we start examining the next two modes in Figs.~\ref{CompareAxial}(j)-(k) and Figs.~\ref{CompareAxial}(f)-(g), we see that they are two nearly degenerate states due to the weak anisotropy generated by the rotating wall, and they are almost related by a rotation by $\pi/2$. Interestingly, as the effective frequency gets smaller in Figs.~\ref{CompareAxial}(b) and (d), we find that the anisotropy forces another mode in Fig.~\ref{CompareAxial}(c) in between these two closely related modes. The remaining mode in Figs.~\ref{CompareAxial}(h) and
(l) is close to the next drumhead mode which has two lines of nodes. This feature is lost with strong enough anisotropy in Fig.~\ref{CompareAxial}(c).  The anisotropy increases because the slower rotating crystals feel the effect of the wall potential more strongly. These first few modes agree well with a continuum elastic theory for the normal modes~\cite{dubin2,john_elastic}.
One interesting feature of all the eigenstates is that the spatial variations are periodic along the circumference of the ion cloud but there is no constraint along the radial direction.
The quantum dynamics of any axial normal mode $\nu$  can be understood as
the collective quantized oscillation with the eigenfrequency $\omega_{z\nu}$ along the $\hat z$ axis with the given snapshot of the eigenvector $b_{j}^{z\nu}$ shown in Fig.~\ref{CompareAxial}.

Due to the presence of the magnetic field, the ion displacement in the
$xy$ plane is chiral arising from the coupling of the dynamics between the $x$ and $y$ components of the same ion. We show that the lower branch of the planar modes is the slow collective renormalized  magnetron modes inherited from the single ion features under $\bf{E}\times \bf{B}$ drift. The higher branch of the planar modes; however, are the fast renormalized cyclotron modes mostly inherited from the the single ion features under the the strong static magnetic field $B_{z}{\hat z}$. The snapshot of the lowest two eigenmodes
in the two branches are shown in Fig.~\ref{ComparePlanar}. The corresponding magnitude of the displacement for a mode $\lambda$ is represented by the length of the arrow and the direction of the arrow represents the displacement of the ion at the time when the snapshot is taken.

However, to understand the character for the planar mode $\lambda$, it is revealing to show the time evolution of the snapshots based on the displacement of the phonon coherent state due to the mode $\lambda$ projected on the ion lattice positions
 $(\langle\delta R_{j}^{x\lambda}(t)\rangle, \langle\delta R_{j}^{y\lambda}(t)\rangle)\propto (|\alpha_{j}^{x\lambda}|\cos\omega_{\lambda}t, |\alpha_{j}^{y\lambda}|\cos\omega_{\lambda}t)$ where the phase $\delta_{\lambda}$ can be chosen to be $-\pi/2$ in Eq.~(\ref{eq:planard}) for our purpose. We provide a movie of this
 in the supplementary material~\cite{supplementary}. Here we summarize what we have observed for the dynamics of the planar modes. Figs.~\ref{ComparePlanar}(b), (f) and (j) show the
 lowest collective cyclotron modes for different effective trapping frequencies. We observe that each ion shows clockwise cyclotron motion around its equilibrium position in the rotating frame.
The second lowest cyclotron modes shown in Figs.~\ref{ComparePlanar} (a), (e), and (i)
also show clockwise cyclotron motions for each ion around its equilibrium but with nonuniform fluctuations on the magnitude of the displacement. The quantum dynamics for the renormalized cyclotron modes are described by the dynamics of the quantized cyclotron orbits.
Since these orbits have much higher energy than the lower branch of the planar modes, the quantized cyclotron orbits have much smaller average radius.

In Figs.~\ref{ComparePlanar}(d), (h), and (l), we show the global shear mode (lowest mode in the lower branch), which is adiabatically connected to the global rotational mode at very weak rotating wall potential, with zero frequency in the rotating frame. The normal mode displacement in principle oscillates back and forth infinitesimally slowly along the elliptical shell of the equilibrium configurations. The snapshots of those modes are taken at different initial conditions and therefore with different orientations of the displacement. In Figs.~\ref{ComparePlanar}(c), (g), and (k),
we observe multiple fragmented shear domains with different signs of circulation. The circulation with one sign of circulation is always suppressed at low energy.

\subsection{Ising spin-spin interactions}
Now that we have described the phonon properties, we are ready to describe the properties of spin-spin interactions when the laser is detuned close to different branches of the phonon modes.

\begin{figure}[htbp!]
  \centering
    \includegraphics[scale=0.4]{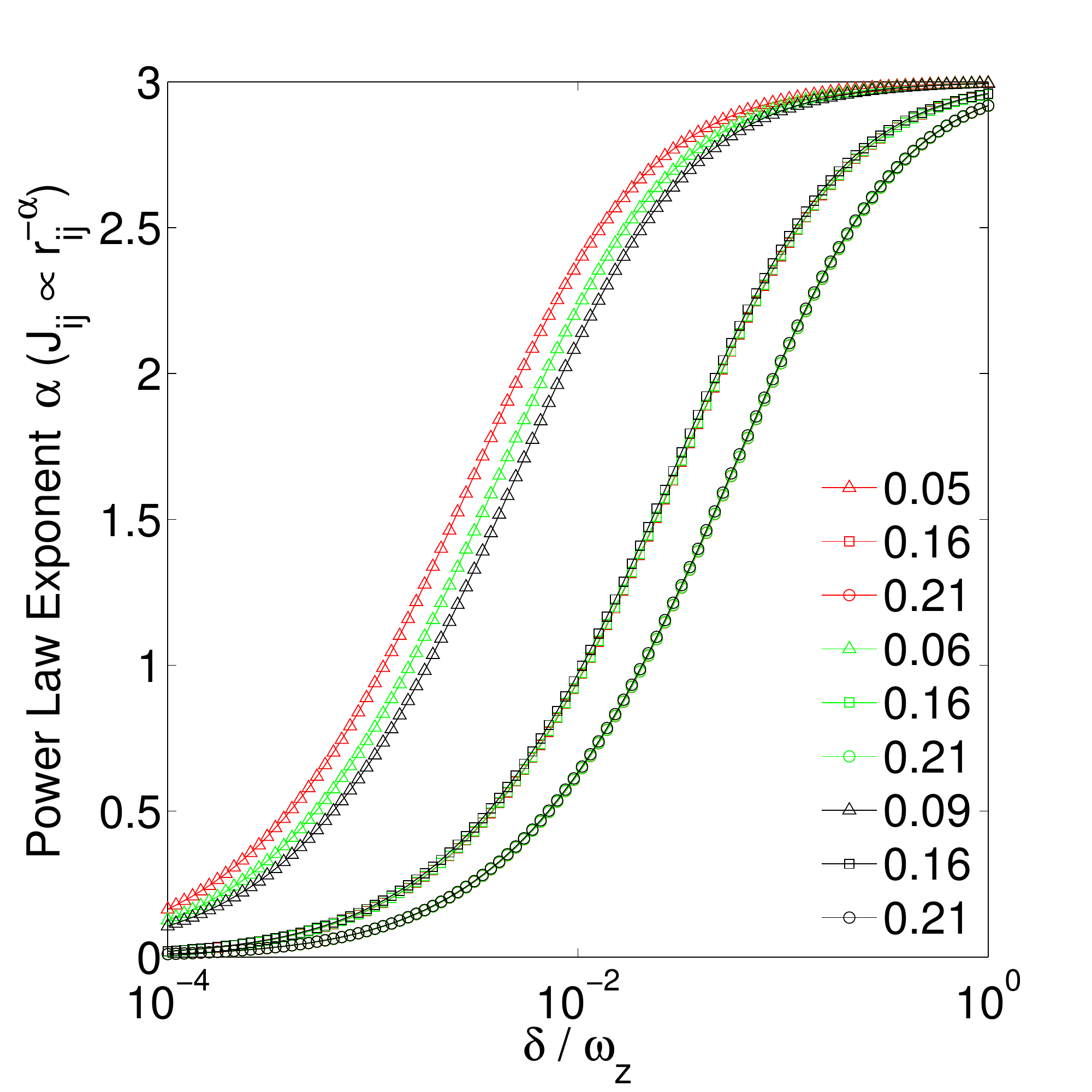}
\caption{(Color online.) Detuning above the axial modes (blue of the COM mode). As in Fig.~2(b) of Ref.~\onlinecite{John}, a detuning $\mu=\delta+\omega_{z}$ above $\omega_z$ yields a power-law-like decay of $J_{ij}\propto r_{ij}^{-\alpha}$ as a
function of the ion distance, $r_{ij}$.
The legend presents the values for the effective trapping frequencies in units of $\omega_{z}$. Here we plot the exponent of the fitted power law as a function of the detuning.}
\label{Above}
\end{figure}

\begin{figure}[htbp!]
  \centering
  \includegraphics[scale=0.45]{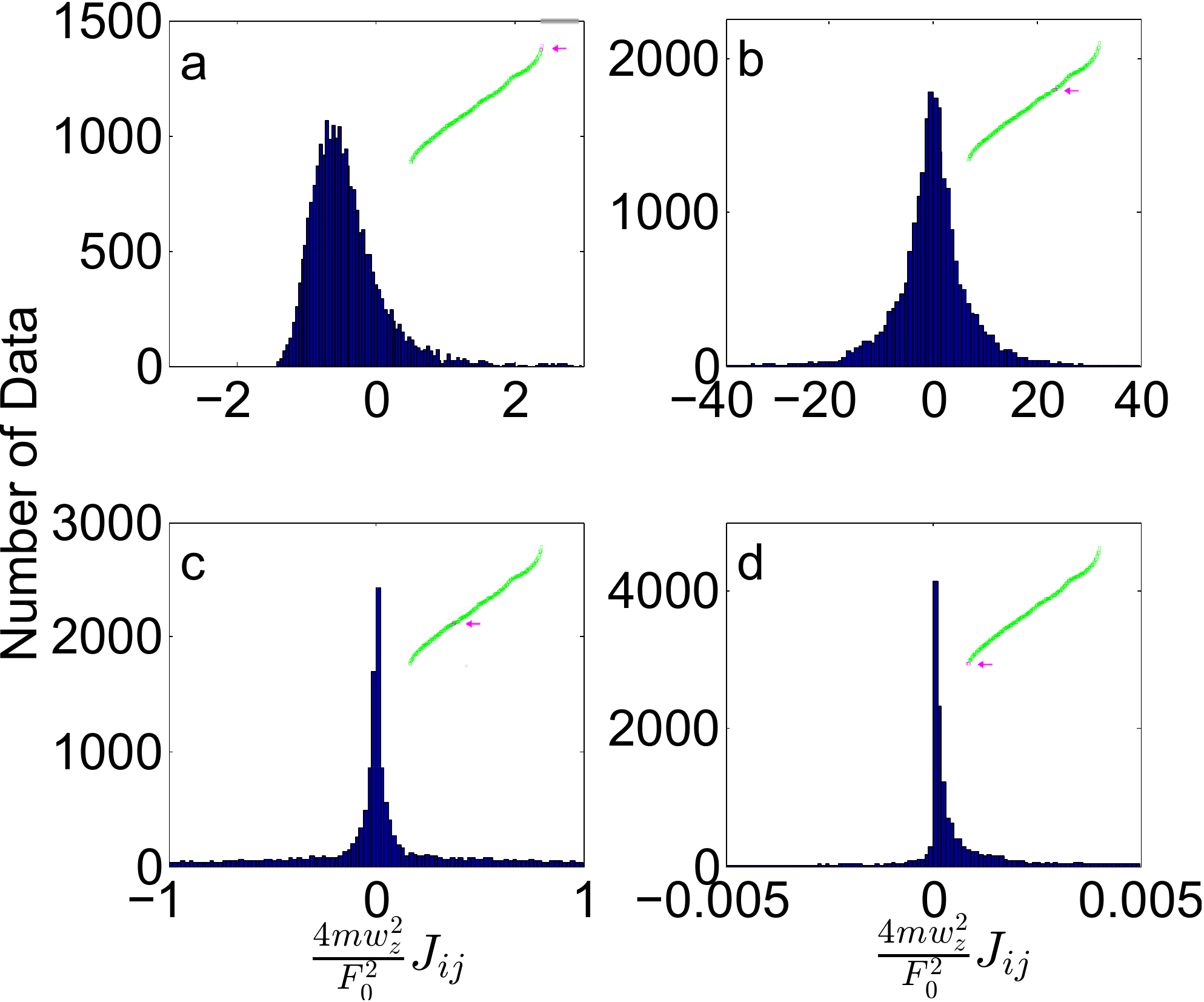}
\caption{(Color online.) Detuning inside the axial mode branch with an effective trapping $\omega_{m}=0.16\omega_{z}$ and the weak rotating wall potential $\omega_{W}=\omega_{l}^{W}=0.06\omega_{z}$ for $N=217$.
(a) Detuning $\delta = 0.00567\omega_z$ below $\omega_z$ exactly between the two highest modes with 99.93\% of the $J_{ij}$ in the plot.
(b) Detuning $\delta = 0.1174\omega_z$ below $\omega_z$ exactly between the $\lfloor 2N/3 \rfloor$ and $\lfloor 2N/3 \rfloor +1$ mode with 99.28\% of the $J_{ij}$ in the plot.
(c) Detuning $\delta = 0.1921\omega_z$ below $\omega_z$ exactly between the $\lfloor N/3 \rfloor$ and $\lfloor N/3 \rfloor +1$ mode with 59.73\% of the $J_{ij}$ in the plot.
(d) Detuning $\delta = 0.2926\omega_z$ below $\omega_z$ exactly between the two lowest modes with 63.65\% of the $J_{ij}$ in the plot. The inserted green curves are generated from Fig.~\ref{Axial}, with the arrow indicating where the detuning lies. }
\label{Inside}
\end{figure}

\begin{figure}[h!]
  \centering
    \includegraphics[scale=0.45]{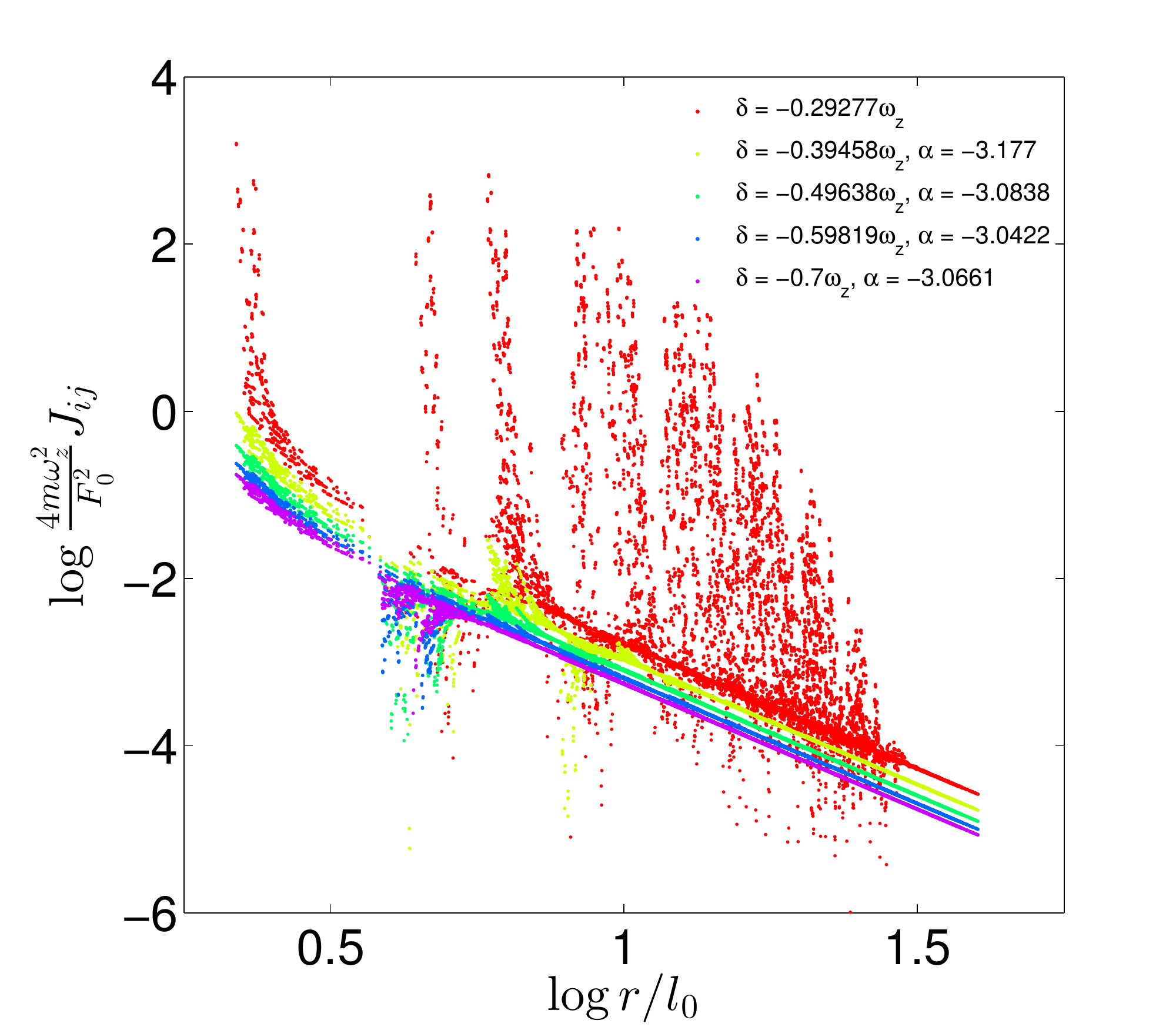}
\caption{Dependence of the spin-spin interaction as a function of the distance $r$ between the ions for a detuning below the axial modes (red of the lowest axial mode).
The effective trapping $\omega_{m}=0.16\omega_{z}$ and the weak rotating wall potential $\omega_{W}=\omega_{l}^{W}=0.06\omega_{z}$ are used.}
\label{Below}
\end{figure}

Let us focus on the axial phonon modes first which have been used
to benchmark the mean spin-spin interactions between hundreds of spins in a Penning trap~\cite{John}.
We calculate the static spin-spin interaction $J_{ij}$ between ions $i$ and $j$ based on Eq.~(\ref{eq:axialspin-spin}).

In Fig.~\ref{Above}, the detunings $\delta=\mu-\omega_{z}$  range from  $0.0001 \omega_z$ to $\omega_{z}$, all detuned to the blue of the COM mode. In this case, all of the interactions are
antiferromagnetic and they decay with a power law depending on the distance between two spins $(J_{ij}\propto r_{ij}^{-\alpha}$). The power law increases from $\alpha=0$, when the detuning is very close to the COM mode to $\alpha=3$, when the detuning is far away from the COM mode. We observe that the case of weak effective trapping
allows for a much more rapid approach to the limit of dipole-dipole interactions ($\alpha = 3$) at large detuning, as shown by the data sets with colored triangles.

We next examine what happens when we detune the spin-dependent optical dipole force to lie in between different axial modes. In this case, the interaction $J_{ij}$ does not reveal well-defined power law behaviors and is frustrated with different signs; therefore,
we report  histograms of the spin-spin interactions.
We choose the following four detunings: the mean of the first and second lowest frequency axial modes,
the mean of the $72$th and $73$th axial modes, the mean of the the $144$th and $145$th axial modes,
and the mean of the first and second highest frequency axial modes.
In Fig.~{\ref{Inside}}, we observe the distribution of the interaction $J_{ij}$ is more symmetric about zero when the detuning $\mu$ is located near the central part of the axial modes such as cases (b) and (c).
In addition, by our more detailed studies, we found that the spin-spin interaction $J_{ij}$ is randomly  disordered with distance but is correlated with the polar angle between ions.
This can be interesting for further studies of disordered spin dynamics with the spin-spin interaction.
However, the interaction $J_{ij}$ becomes asymmetric when the detuning $\mu$ is located elsewhere.
We also notice that the fluctuations of $J_{ij}$ are larger when the detuning is closer to the high frequency portion of the axial modes.

Below the axial modes, the detunings are chosen with an equal spacing from just below the lowest frequency axial mode,
to $0.3\omega_z$ which is sufficiently far from the lower branch planar modes \emph{for the effective trapping $\omega_m$}.
For large detuning $|\delta|$, we still observe the majority of the spin-spin interaction $J_{ij}$ follows the power law scaling $J_{ij}\propto 1/r_{ij}^{\alpha}$ except for the case with very small detuning such as the case with $\delta = -0.29277\omega_{z}$ in Fig.~\ref{Below}.

\begin{figure}[h!]
  \centering
    \includegraphics[scale=0.28]{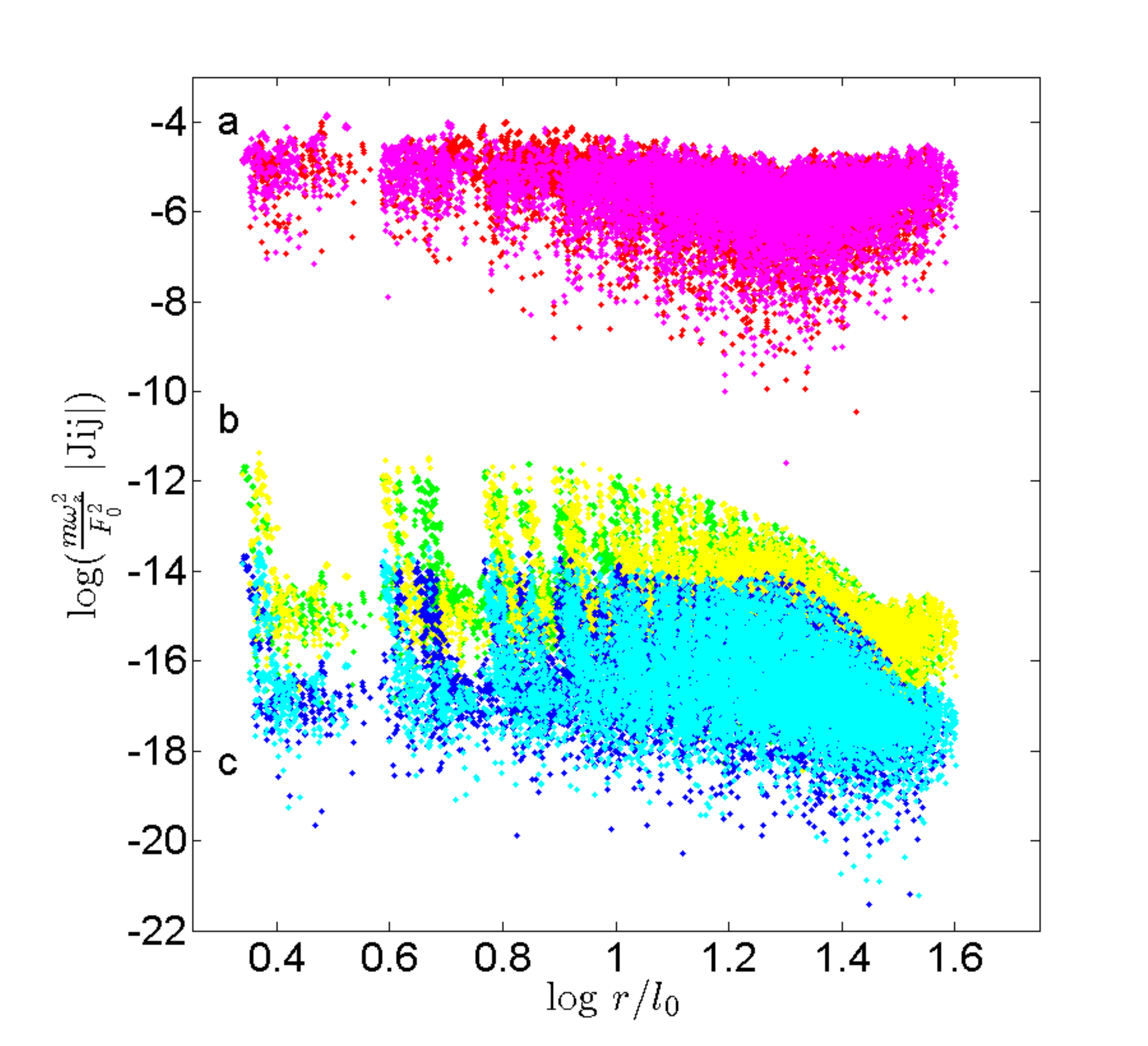}
\caption{Dependence of planar spin-spin interaction as a function of the distance $r$ between ions for various detunings:
(a) $\mu = 1\e{-5}\omega_z$;(b) $\mu = \omega_z$;(c) $\mu = 10\omega_z$. The darker colors illustrate positive $J_{ij}$ and the lighter colors
illustrate negative $J_{ij}$.
The effective trapping $\omega_{m}=0.16\omega_{z}$ and the weak rotating potential $\omega_{W}=\omega_{l}^{W}=0.06\omega_{z}$ are used.}
\label{PJijR}
\end{figure}

Similar to calculating the spin-spin interaction $J_{ij}$ for the axial modes, we use Eq. (\ref{eq:PlanarJij}) to calculate the $J_{ij}$ for the planar modes.
In Fig.~\ref{PJijR}, we present the planar spin-spin interaction $J_{ij}$ as a function of the distance $r$ between ions. We show that the couplings $J_{ij}$ do not follow a well-defined power law scaling, but the distribution better resembles the symmetric histogram in Fig. \ref{Inside}(b).
 Detuning the lasers near the lowest lower branch planar mode at $\mu = 1\e{-5}\omega_z$, in between the lower and upper branch planar modes at
 $\mu = \omega_z$, and above the upper branch planar modes at $\mu = 10\omega_z$, corresponds to Figs.~\ref{PJijR}(a), (b), and (c), respectively. The darker colors (red, green, blue)
 represent positive $J_{ij}$ and the lighter colors (magenta, yellow, cyan) represent negative $J_{ij}$ values for the same parameter set. One can observe all $J_{ij}$ are almost equally distributed among positive and negative values for all the cases
 without showing any clear signature of power law behavior.

\begin{figure}[h!]
  \centering
    \includegraphics[scale=0.33]{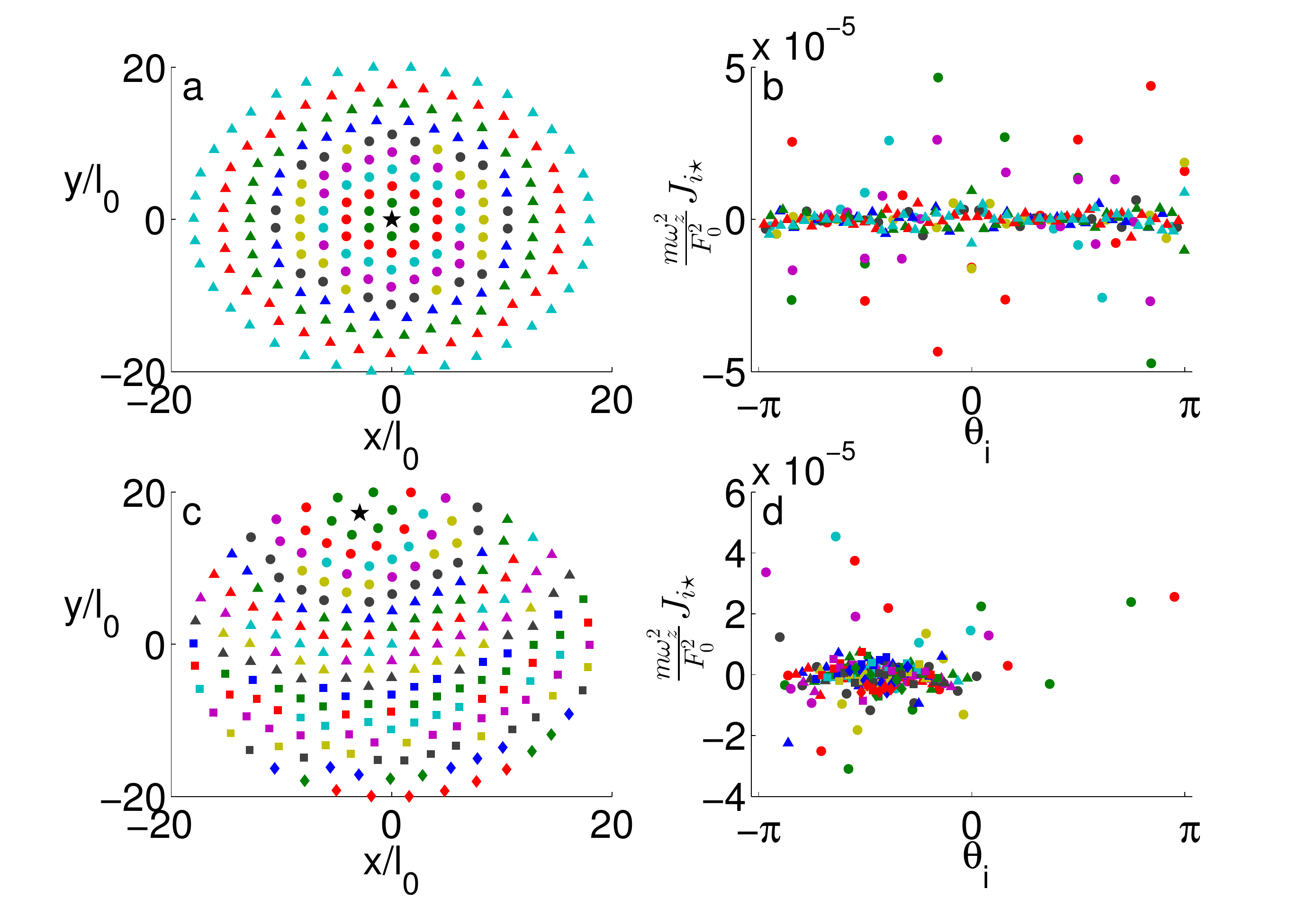}
\caption{Angular correlation of $J_{ij}$ with $\mu = 1\e{-5}\omega_z$. (a) and (c) The subshell definition for the ion at the center
and for an outer ion. (b) and (d) The spin-spin interaction $J_{ij}$ of these subshells as a function of radius r.
The effective trapping $\omega_{m}=0.16\omega_{z}$ and the weak rotating wall potential $\omega_{W}=\omega_{l}^{W}=0.06\omega_{z}$ are used.}
\label{PJijCorr}
\end{figure}

In Fig.~\ref{PJijCorr}, we examine whether there is any peculiar spatial dependence to the $J_{ij}$ patterns for the case with $\mu = 1\e{-5}\omega_z$. In Figs.~\ref{PJijCorr}(a) and (c),
we select the center and an outer ion as the reference point respectively (shown with the black star $\star$) and calculate the interaction $J_{i\star}$ between each ion $i$  with respect to the reference ion.
In order to study the dependence of $J_{i\star}$ on the ion distance $r$, we define the subshells of the crystal
as ions that have a similar radius from the reference ion. In Figs.~\ref{PJijCorr}(a) and (c), ions are in the same subshell if they are illustrated with the same shape and color.
From this new origin, we define  $\theta_{i}$ as the polar angle with  respect to the reference ion and plot the $J_{i\star}$ from each subshell as a function of $\theta_{i}$ as
shown in Figs.~\ref{PJijCorr}(b) and (d). Each of these figures show that the magnitude $|J_{i\star}|$ in each shell increases with radius as shown by the largest amplitude of the oscillations for the ions in the same shell (data points with the same color data) and $J_{i\star}$ alternate in sign  at a certain fixed orientation $\theta_{i}$.
In addition, the interaction $J_{i\star}$ alternates within the same subshell, which is very different from the axial modes. These oscillatory correlations appear to exist for other detunings such as $\mu = \omega_z$ or $\mu = 10\omega_z$ as well.
Noticeably, as shown in Fig.~17(c), the interaction $J_{i\star}$ is stronger for ions situated along the orientation $\theta_{i}=\pi$ opposite to the orientation of  the laser wavevector $\delta k_{x}$($\theta_{i}=0$) due to the symmetry breaking of the laser beam.
So, in general, the spin model realized by coupling to the planar modes creates frustration due to randomness and long-range interaction for spins, which can be potentially be used as a platform to explore spin-glass physics, especially since hundreds of ions
are accessible in the Penning trap systems.

\section{Conclusion and discussion}

In this work, we have presented a thorough treatment of the theoretical background for further development of cold ion trap computation in a Penning trap, bringing the theory to a similar level to what has been available for linear Paul trap simulators for years~\cite{spin-spin-interactions,James}. In particular, we have shown how to determine the equilibrium positions efficiently, how to find the phonon modes for motion about the equilibrium positions, and how to couple the phonon motion to spin-dependent forces to determine the effective spin exchange between two different ions. Our work focused on both axial and planar modes.  The latter are more involved because they have velocity-dependent forces in the rotating frame due to the Coriolis force and Lorentz force, and require more complicated methods to perform the normal-mode analysis.

Our results show, that in some cases the system is described by relatively simple Ising models, which can have a transverse magnetic field added in order to investigate adiabatic state preparation. But, in many cases, the spin-exchange patterns are quite complex, and will require significant analysis to understand their behavior. We were forced to restrict ourselves to examine just a handful of different cases in this work, because the parameter space is so huge.  There may very well be a number of interesting new results that can be found by applying our approach to different situations than what we examined. In any case, this formalism will be quite important in analyzing the next generation of experiments on Penning-trap-based cold ion computation, especially for cases where one maps onto Ising-like models.

Fully understanding actual experiments will, of course, require much deeper analysis than what was given here. There are intrinsic errors with respect to the target spin Hamiltonian, noise due to external environments, and spontaneous emission effects, which may deteriorate
the quantum simulation. For example, in the presence of a transverse magnetic field in an adiabatic state preparation protocol, it may not be possible to completely keep the phonon degrees of freedom as only virtual excitations during the simulation~\cite{Phonons}.
In addition, there are errors in the initialization of quantum spin states and of the phonon states. For example, when the crystal is not sufficiently cold, there will be a background of phonons present prior to the start of the simulation. Those can modify the fidelity of the quantum simulation and understanding how this works requires future study.

\section{Acknowledgements}
 The authors are thankful to John Bollinger, Brian Sawyer, Joe Britton, and Michael Biercuk for valuable discussions. This work was supported under ARO grant number W911NF0710576 with funds from the DARPA OLE Program. J. K. F. also acknowledges the McDevitt bequest at Georgetown University. A. C. K. also acknowledges support of the National Science Foundation under grant number DMR-1004268 during the initial stages of this work in the summer of 2011.

\appendix
\section{Details for solving the quadratic eigenvalue problem}

We discuss some of the general properties of the quadratic eigenvalue problem here, since this problem is not too well known within the physics community.  Our approach is closely related to the work of Baiko~\cite{baiko}, but generalizes his work to the case we need for the Penning trap planar phonons. For a more mathematical discussion, see Ref.~\onlinecite{QEP}.

The planar phonon  Hamiltonian (in second quantized form) is given by
$H_{ph}^{P}=\sum_{\lambda=1}^{2N}
\hbar\omega_{\lambda}(a_{\lambda}^{\dagger}a_{\lambda}+\frac{1}{2})$
with the phonon frequencies all positive ($\omega_{\lambda}>0$).
In terms of the eigenvectors $\alpha_{\lambda}^{\nu}$ discussed in Sec. II. C, we
can solve the collective mode problem by solving the following QEP and choosing solutions with positive eigenvalues $\omega_{\lambda}>0$
\begin{equation}
\Big(m\omega_{\lambda}^2\delta_{\nu\nu'}+i\omega_{\lambda}T_{\nu\nu'}
-m{\omega_{0}^{\nu}}^2\delta_{\nu\nu'}\Big)\alpha_{\lambda}^{\nu'}=0.
\label{app:quad}
\end{equation}
Here $\lambda$ is the label for the different solutions and $\nu$ is the matrix/vector index which is summed over.  Hence, this is a $2N\times 2N$ quadratic eigenvalue problem, with eigenvectors $\alpha^\nu_\lambda$.  There are $4N$ eigenvalues, and hence $4N$ associated eigenvectors. Because the QEP is complex, it turns out that the eigenvectors are not orthogonal, and hence the behavior is different from a conventional linear eigenvalue problem.  We will derive the relations that the eigenvectors do satisfy by mapping the QEP onto a linear eigenvalue problem and using the well-known properties of those solutions (this procedure is called linearizing the QEP).

The QEP can be mapped onto a linear hermitian eigenvalue problem as follows:
\begin{equation}
\left(
  \begin{array}{cc}
    -\frac{i}{m}T_{\nu\nu'} & \omega_{0}^{\nu}\delta_{\nu\nu'} \\
    \omega_{0}^{\nu}\delta_{\nu\nu'} & 0\\
  \end{array}
\right)\left(
         \begin{array}{c}
           \omega_{\lambda}\alpha_{\lambda}^{\nu'} \\
           \omega_{0}^{\nu'}\alpha_{\lambda}^{\nu'} \\
         \end{array}
       \right)
=\omega_{\lambda}\left(
                   \begin{array}{c}
                     \omega_{\lambda}\alpha_{\lambda}^{\nu} \\
                     \omega_{0}^{\nu}\alpha_{\lambda}^{\nu} \\
                   \end{array}
                 \right).
\label{eq: linearfromqep}
\end{equation}
The $4N\times 4N$ matrix is hermitian since $T$ is real and antisymmetric.  Note that the QEP, and hence the corresponding linear eigenvalue problem, both have positive and negative eigenvalues; for the physical solutions we want, we will restrict to positive eigenvalues only.

It may seem odd that the eigenvalue appears as part of the eigenvector for this solution, which might bring some confusion about how one solves such a problem.  If we let $M_{\nu\nu'}$ denote the matrix on the left hand side of Eq.~(\ref{eq: linearfromqep}), and let $x_\lambda^\nu$ denote the eigenvector, then the equation is a conventional eigenvalue problem $M_{\nu\nu'}x_\lambda^{\nu'}=\omega_\lambda x_\lambda^\nu$.  After solving this conventional problem, the $\alpha_\lambda^\nu$ vectors are extracted from the explicit forms given in Eq.~(\ref{eq: linearfromqep}). One might be concerned about whether the eigenvectors can always be written in this form, but this always follows from the specific form of the matrix $M$.

By taking the complex conjugate of Eq.~(\ref{app:quad}), we find that if $\alpha_\lambda^\nu$ is the $\lambda$th eigenvector of the QEP with eigenvalue $\omega_\lambda$, then $\alpha_\lambda^{\nu*}$ is the eigenvector of the QEP with eigenvalue $-\omega_\lambda$. Hence the eigenvalues come in $\pm$ pairs, with corresponding eigenvectors of the linear eigenvalue problem $(\omega_{\lambda}\alpha_{\lambda}^{\nu} \ \omega_{0}^{\nu}\alpha_{\lambda}^{\nu})^{T}$ with positive eigenvalue $\omega_{\lambda}>0$,
and $(-\omega_{\lambda}\alpha_{\lambda}^{\nu *} \
\omega_{0}^{\nu}\alpha_{\lambda}^{\nu*})^T$
 with negative eigenvalue $-\omega_{\lambda}<0$ as long as the eigenvalues
 $\omega_{\lambda}$ and $\omega_{0}^\nu$ are real.
As a consequence, there are $4N$ eigensolutions (as expected for a $4N\times 4N$ linear eigenvalue problem) and half of them have positive eigenvalues $\omega_{\lambda}>0$.  We organize them as follows with the positive eigenvalues $(\omega_\lambda>0$) in the first $2N$ entries and the negative ones $(-\omega_\lambda <0)$ in the second $2N$:
\begin{eqnarray}
x_\lambda^\nu&=&\Biggr [  \begin{array}{l l}
                     \omega_{\lambda}\alpha_{\lambda}^{\nu}/c_\lambda &{\rm for}\quad 1\le\nu\le 2N\\
                     \omega_{0}^{\nu-2N}\alpha_{\lambda}^{\nu-2N}/c_\lambda  &{\rm for}\quad 2N+1\le\nu\le 4N\\
                   \end{array}\nonumber\\
                  &\quad&{\rm for}\quad 1\le\lambda\le2N \nonumber \\
x_\lambda^\nu&=&\Biggr [  \begin{array}{l l}
                     -\omega_{\lambda}\alpha_{\lambda}^{\nu*}/c_\lambda &{\rm for}\quad 1\le\nu\le 2N\\
                     \omega_{0}^{\nu-2N}\alpha_{\lambda}^{\nu-2N*}/c_\lambda  &{\rm for}\quad 2N+1\le\nu\le 4N\\
                   \end{array}\nonumber\\
                  &\quad&{\rm for}\quad 2N+1\le\lambda\le 4N.
\label{eq: x_def}
\end{eqnarray}
The eigenvectors $x_\lambda$ form a complete and orthonormal set.  The constants $c_\lambda=c_{\lambda-2N}$ are chosen to provide the needed normalization of the $\alpha_\lambda$ eigenvectors, which are not orthogonal, and are not normalized to unit norm, as described below.

Because the linear eigenvalue problem is hermitian, its eigenvectors are orthonormal, implying $\sum_{\nu}x_\lambda^{\nu *}x_{\lambda'}^\nu=\delta_{\lambda\lambda'}$.  If we choose $\lambda$ and $\lambda'$ to both lie between 1 and $2N$, we find
\begin{equation}
\sum_{\nu=1}^{2N}(\omega_\lambda\omega_{\lambda'}+\omega_0^{\nu 2})\alpha_\lambda^{\nu *}\alpha_{\lambda'}^\nu=c_\lambda\delta_{\lambda\lambda'}.
\label{eq: alpha_orthog}
\end{equation}
We choose $c_\lambda=\omega_\lambda/\hbar m$ in order to satisfy the normalization condition for the collective mode raising and lowering operators $[a_\lambda,a^\dagger_{\lambda'}]=\delta_{\lambda\lambda'}$ in Eq.~(\ref{eq:normalization}). We will choose the exact same positive value for $c_\lambda$ for the corresponding negative eigenvalue solutions as well ($c_\lambda=c_{\lambda-2N}$).

We can derive three more relations between the $\alpha$ eigenvectors by using the completeness of the $x$ eigenvectors $\sum_{\lambda=1}^{4N}x_\lambda^{\nu *}x_\lambda^{\nu'}=\delta_{\nu\nu'}$.  We start with $\nu$ and $\nu'$ both restricted between 1 and $2N$.  This gives
\begin{equation}
\sum_{\lambda:\omega_\lambda>0}\omega_\lambda(\alpha_\lambda^{\nu *}\alpha_\lambda^{\nu'}+\alpha_\lambda^{\nu}\alpha_\lambda^{\nu ' *})=\hbar m\delta_{\nu\nu'}.
\label{eq: alpha_comp1}
\end{equation}
Next, taking $1\le\nu\le 2N$ and $2N+1\le\nu'\le 4N$ yields
\begin{equation}
\sum_{\lambda:\omega_\lambda>0}(\alpha_\lambda^{\nu *}\alpha_\lambda^{\nu'}-\alpha_\lambda^\nu\alpha_\lambda^{\nu'*})=0.
\label{eq: alpha_comp2}
\end{equation}
Finally, taking both $\nu$ and $\nu'$ to be larger than $2N$ gives
\begin{equation}
\sum_{\lambda:\omega_\lambda>0}\frac{1}{\omega_\lambda}(\alpha_\lambda^{\nu *}\alpha_\lambda^{\nu'}+\alpha_\lambda^\nu\alpha_\lambda^{\nu' *})=\frac{1}{\hbar m\omega_0^{\nu 2}}\delta_{\nu\nu'}.
\label{eq: alpha_comp3}
\end{equation}
In all of these relations, the second terms in the parentheses arise from the negative eigenvalue solutions of the QEP.


\begin{thebibliography}{99}
\bibitem{feynman}R. P. Feynman, Int. J. Theor.
Phys. {\bf 21}, 467 (1981).
\bibitem{two-ion}
A. Friedenauer, H. Schmitz, J. T. Glueckert, D. Porras, and T.
Schaetz, Nat. Phys. {\bf 4}, 757 (2008).
\bibitem{Kim1}
K. Kim, M.-S. Chang, R. Islam, S. Korenblit, L.-M. Duan, and
C. Monroe, Phys. Rev. Lett. {\bf 103}, 120502 (2009).
\bibitem{Kim2} K. Kim, M.-S. Chang, S. Korenblit, R. Islam, E. E. Edwards, J. K. Freericks, G.-D. Lin, L.-M. Duan, and C. Monroe, Nature, {\bf 465}, 590 (2010).
\bibitem{Edwards} E. E. Edwards, S. Korenblit, K. Kim, R. Islam, M.-S. Chang, J. K. Freericks, G.-D. Lin, L.-M. Duan, C. Monroe,  Phys. Rev. B {\bf  82}, 060412(R) (2010)
\bibitem{Islam} R. Islam, E. E. Edwards, K. Kim, S. Korenblit, C. Noh, H. Carmichael, G.-D.Lin, L.-M. Duan, C.-C. Joseph Wang, J. K. Freericks, and C. Monroe, Nature Commun. {\bf 2}, 1374 (2011).
\bibitem{Markus} J. Simon, W. S. Bakr, R. Ma, M. E. Tai, P. M. Preiss and Markus Greiner, Nature {\bf 472}, 307 (2011).
\bibitem{John} J. W. Britton, B. C. Sawyer, A. C. Keith, C.-C. Joseph Wang, J. K. Freericks, M. J. Biercuk, H. Uys, and J. J. Bollinger, Nature {\bf 484}, 489 (2012).

\bibitem{Marcos} M. Rigol, V. Dunjko, and M. Olshanii, Nature {\bf 452}, 854 (2008).

\bibitem{dynamic QS}S. Trotzky, Y-A. Chen, A. Flesch, I. P. McCulloch, U. Schollw\"{o}ck, J. Eisert and I. Bloch, Nat. Phys. {\bf 8},
 325 (2012).


\bibitem{Santos} L. F. Santos, A. Polkovnikov, and Marcos Rigol, Phys. Rev. Lett. {\bf 107}, 040601 (2011).

\bibitem{kastner} M. Kastner, Phys. Rev. Lett. {\bf 106}, 130601 (2011).

\bibitem{spin-spin-interactions}D. Porras and J. I. Cirac, Phys. Rev. Lett. {\bf 92}, 207901 (2004).

\bibitem{James}D. F. V. James, Appl. Phys. B {\bf 66}, 181 (1998); C. Marquet, F. Schmidt-Kaler and D. F. V. James, Appl. Phys. B {\bf 76}, 199 (2003).

\bibitem{paultrap_2d} H. Kaufmann, S. Ulm, G. Jacob, U. Poschinger, H. Landa, A. Retzker, M.B. Plenio, F. Schmidt-Kaler, arxiv:1208.4040 (2012).

\bibitem{Chinese}
P. Zou, J. Xu, W. Song, and S.-L. Zhu, Phys. Lett. A {\bf 374}, 1425 (2010).

\bibitem{Jake}
J. D. Baltrusch, A. Negretti, J. M. Taylor, and T. Calarco, Phys. Rev. A {\bf 83}, 042319 (2011).

\bibitem{John2} M. J. Biercuk, H. Uys, A. P. Van Devender, N. Shiga, W. M. Itano, J. J. Bollinger, Quant. Info. and Comp. {\bf 9}, 920 (2009).

\bibitem{rotatingwall} T. Hasegawa, M. J. Jensen, and J. J. Bollinger, Phys. Rev. A {\bf 71}, 023406 (2005).

\bibitem{dubin1} D. H. E. Dubin, Phys. Rev. Lett. {\bf 71}, 2753 (1993).

\bibitem{Brian} B. C. Sawyer, J. W. Britton, A. C. Keith, C.-C. Joseph Wang, J. K. Freericks, H. Uys, M. J. Biercuk, J. J. Bollinger,
    Phys. Rev. Lett. {\bf 108}, 213003 (2012).

\bibitem{John3} X.-P. Huang, J. J. Bollinger, T. B. Mitchell, and Wayne M. Itano, Phys. Rev. Lett. {\bf 80}, 73 (1998).

\bibitem{dubin_oneil}
D. H. E. Dubin and T. M. O'Neil, Rev. Mod. Phys. {\bf 71}, 87 (1999).

\bibitem{japanese}
H. Fukuyama, Sol. St. Commun. {\bf 17}, 1323 (1975); T. Nagai and H. Fukuyama, J. Phys. Soc. Japan {\bf 51}, 3431 (1982); T. Nagai and H. Fukuyama, J. Phys. Soc. Japan {\bf 52}, 44 (1983).

\bibitem{russian_old}
N. A. Usov, Yu. B. Grebenshchikov, and F. R. Ulinich, Zh. Eksp. Teor. Fiz. {\bf 78}, 296 (1980) [Sov. Phys. JETP {\bf 51}, 148 (1980)].

\bibitem{chen_thesis}
Shi-Jie Chen, ``Temperature equilibration and many-particle adiabatic invariants'', University of California, San Diego, Ph.D. Thesis (1994).

\bibitem{baiko}
D. A. Baiko, Phys. Rev. E {\bf 80}, 046405 (2009).

\bibitem{QEP} F. Tisseur and K. Meerbergen, SIAM Rev. {\bf 43}, 235 (2001).

\bibitem{PJ} P. J. Lee, K.-A. Brickman, L. Deslauriers, P. C. Haljan, L.-M. Duan, and C. Monroe, J. Opt. B {\bf 7}, 371 (2005).

\bibitem{Phonons} C.-C. Joseph Wang and J. K. Freericks,
Phy. Rev. A  {\bf 86}, 032329 (2012).

\bibitem{Partial}
$ \exp[{ikr\cos \theta }]=\sum_{l}i^{l}(2l+1)j_{l}(kr)P_{l}(\cos \theta)$ where $\theta\equiv -\Omega t+\phi_0$.

\bibitem{peeters}
V. A. Schweigert and F. M. Peeters, Phys. Rev. B {\bf 51}, 7700 (1995).

\bibitem{dubin2} D. H. E. Dubin, Phys. Rev. Lett. {\bf 66}, 2076 (1991).

\bibitem{john_elastic}
J. J. Bollinger, D. J. Heinzen, F. L. Moore, W. M. Itano, D. J. Wineland, and D. H. E. Dubin, Phys. Rev. A {\bf 48}, 525 (1993).

\bibitem{supplementary} The supplementary movie can be found with the materials on the APS website.


\end{thebibliography}
\end{document}